\documentclass[aps,prl,twocolumn,amsmath,amssymb,superscriptaddress,10pt]{revtex4-2}
\usepackage{fix-cm}
\usepackage{array}
\usepackage{bm}
\usepackage{color}
\usepackage{float}
\usepackage{graphicx}
\usepackage[caption=false]{subfig}

\usepackage{newtxtext}
\usepackage{newtxmath}

\usepackage{enumitem}
\usepackage{gensymb}
\usepackage{natbib}
\usepackage{soul}
\usepackage{lipsum}
\usepackage{hyperref}
\hypersetup{colorlinks=true,allcolors=blue}

\newcommand{\kag}{kagome}
\newcommand{\hs}[2]{$#1$H$#2$S}

\newcommand{\rord}{$\sqrt{3}\times\sqrt{3}$}

\makeatletter
\newcommand\footnoteref[1]{\protected@xdef\@thefnmark{\ref{#1}}\@footnotemark}
\makeatother

\begin{document}

\title{Kinetic kagome magnetism: from self-trapping RVB polarons to semiclassical correlations}

\author{Yufei Pei}
\affiliation{TCM Group, Cavendish Laboratory, JJ Thomson Avenue, Cambridge CB3 0US, United Kingdom}
\affiliation{Max Planck Institute for the Physics of Complex Systems, 01187 Dresden, Germany}

\author{Shuai A. Chen}
\affiliation{Max Planck Institute for the Physics of Complex Systems, 01187 Dresden, Germany}

\author{Claudio Castelnovo}
\affiliation{TCM Group, Cavendish Laboratory, JJ Thomson Avenue, Cambridge CB3 0US, United Kingdom}

\author{Roderich Moessner}
\affiliation{Max Planck Institute for the Physics of Complex Systems, 01187 Dresden, Germany}

\begin{abstract}
To gain deeper insight into the role of hole kinetics in determining magnetism in highly frustrated doped Mott insulators,
we consider the single-hole counter-Nagaoka problem on the {\kag} lattice, using magnetization as a tuning parameter.
Near full polarization, a doped hole delocalizes upon binding reversed spins in a pattern of singlet bonds which we term resonating-valence-bond (RVB) polaron. These RVB polarons can have extremely small effective bandwidths, and hence exhibit self-trapping. 
By tuning the spin polarization, we track the evolution of these states toward the unpolarized sector, where we observe the emergence of $\sqrt{3}\times\sqrt{3}$ antiferromagnetic correlation reminiscent of the classical Potts and Heisenberg models on the {\kag} lattice. 
These results provide a framework to understand how RVB physics at short
scales evolves into conventional magnetic correlations at long scales.
\end{abstract}
\maketitle
%
%
The interplay between itinerant particles and magnetism is a constitutive problem of strongly correlated electron physics. 
Pioneering work by Nagaoka~\cite{thouless1965exchange,nagaoka1966ferromagnetism,tasaki1989extension} showed that (ferro)magnetic correlations can originate from the kinetic energy of a sufficiently low density of dopants in infinite-$U$ Hubbard systems, where virtual exchange processes
are absent. 
When said kinetic energy is frustrated, it was later shown that antiferromagnetic correlations can also appear~\cite{Mielke1992,Brandt1992,Tasaki1993,Penc1999}. 

Seminal work by Haerter and Shastry~\cite{haerter2005kinetic} set the stage for the study of kinetic antiferromagnetism in many-body systems (see also Ref.~\onlinecite{Iordanskii1980} for an earlier study). 
Research efforts to date focused primarily on the two-dimensional triangular lattice, where kinetic frustration leads to $120^\circ$ antiferromagnetic order~\cite{sposetti2014classical}. 
This behaviour is in notable contrast with studies on chains~\cite{Glittum2026}, Husimi cactus graphs~\cite{kim2023exact}, and non-planar lattices~\cite{Tasaki1993,glittum2025resonant}, where instead the magnetic correlations can be understood in terms of valence bond singlets. 
Interest and research activities in these systems has seen a substantive resurgence of late, also thanks to advances in computational approaches -- in particular density-matrix renormalization group~\cite{jiang2017holon,zhu2022doped,chen2022proposal, morera2023high} -- and experimental progress in moiré materials and cold-atomic systems~\cite{tang2020simulation,ciorciaro2023kinetic,xu2023frustration,lebrat2024observation, prichard2024directly}. 

In this work we study the prototypical example of such Hamiltonians -- a kinetically-frustrated singly-doped, infinite-$U$ Hubbard model -- on the {\kag} lattice, the archetypical highly frustrated lattice in two dimensions. By progressing through different magnetization sectors, we are able to show how valence bond singlet physics underpins the behavior of the system close to magnetic saturation, leading to the formation of what we dub RVB polarons. As the number of reversed spins increases, resonant processes between overlapping valence bond patterns gradually give rise to magnetic correlations that are consistent with semiclassical $\sqrt{3}\times\sqrt{3}$ order in the unpolarized sector.
Our model provides an important conceptual bridge between singlet physics and conventional antiferromagnetism borne out of kinetic frustration. 

Interestingly, we find that the polarons formed by holes and reversed spins interplay in a non-trivial way with the localization properties typical of the {\kag} lattice~\cite{Mielke1991,Mielke1992ka}. 
Indeed, kinetic frustration arises when the {\kag} flat band happens to be lowest in energy for a hole-doped fully polarized electron system close to half-filling. 
As such, holes in a fully polarized system occupy compact localized states~\cite{Mielke1992ka} at low energy. 
Upon adding few reversed spins, polarons form and delocalize, albeit with a band mass $100$ times larger than that of an unfrustrated tight-binding particle on the same lattice.
Once the number of reversed spins is large enough for the valence bond pattern to resonate around closed loops, the polarons appear to re-localize, in that  the best of our numerics shows a further and sudden increase in band mass of at least a factor of $10^6$. 
Eventually, in the unpolarized limit, the large density of reversed spins succeeds to relieve the kinetic frustration and delocalized, dispersive holes are recovered.

Our work uncovers an intricate interplay between single-particle flat bands and cooperative `polaronic' self-trapping physics, as well as between resonating-valence-bond physics and semiclassical magnetic correlations. 
In this sense, the counter-Nagaoka effect on the {\kag} lattice presents a variable and dense fabric woven out of various strands of correlated and topological electron physics, whose joint origins lie in the study of high-temperature superconductivity and frustrated magnetism. 
%
%

\begin{figure}
    \centering
    \subfloat[]{%
        \raisebox{1.5em}{
            
        \includegraphics[width=0.35\columnwidth]{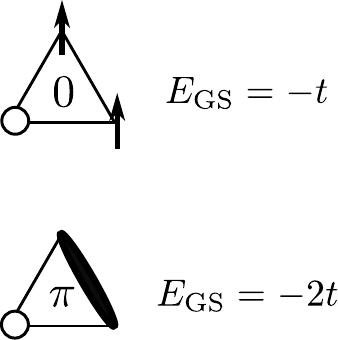}
        \label{fig:1a}
        }
    }
    \hfill
    \subfloat[]{%
        \includegraphics[width=0.61\columnwidth]{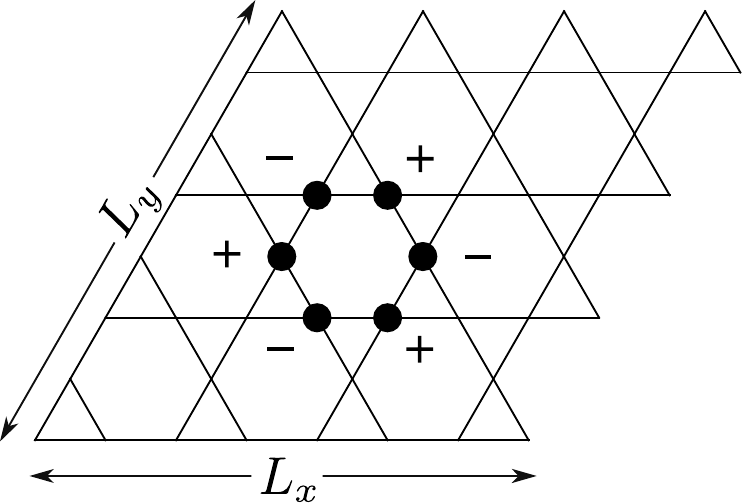}
        \label{fig:1b}
    }
    \caption{\label{fig:1}(a) Kinetic magnetism in the presence of frustration favors the formation of valence bond singlets. 
    (b) Kagome lattice geometry used in our simulations, for system size $N=3 L_x L_y$. We also show an example of a compact localized state around a single plaquette~\cite{Mielke1992ka}. 
    }
\end{figure}

\emph{Model} --- 
We consider the infinite-$U$ Hubbard model on the {\kag} lattice, 
\begin{equation}\label{eq:hamiltonian}
    \mathcal{H}=-t\sum_{\langle ij\rangle,\, \sigma}c_{i\sigma}^\dagger c_{j\sigma}\,,
\end{equation}
where $\langle ij \rangle$ enumerate nearest neighbor sites, and $c^\dagger_{i\sigma}$ ($c_{i\sigma}$) are creation (annihilation) operators for electrons with spin $\sigma\in \{\downarrow, \uparrow\}$ at site $i$, 
subject to projecting out the states with double occupancy. 
Kinetic frustration arises when $t>0$ (set hereafter to $t=1$ as our unit of energy). This can be seen straightforwardly by solving the $2$-electron problem on a single triangle, as illustrated in Fig.~\ref{fig:1a}. Ferromagnetic spins have lowest energy $E_0=-t$ due to destructive interference of the hole motion around the triangle; whereas antiferromagnetic spins can form a singlet and effectively thread a $\pi$ flux through the triangle, relieving the frustration and allowing the hole to delocalize and lower the energy to $E_0=-2t$ (bottom of an unfrustrated 1D tight-binding band). 

In the following we study this system using exact diagonalization (ED: Lanczos~\cite{lanczos1950iteration}, XDiag~\cite{xdiag}) and density matrix renormalization group (DMRG: TenPy~\cite{tenpy2024}), in the YC-ones geometry shown in Fig.~\ref{fig:1b} (see also EM). 
%
%

\emph{RVB polarons and delocalization} --- 
The fully polarized case reduces to a well-known spinless fermionic tight-binding model, where the bottom band is exactly flat (energy $-2$) and the states~\footnote{With subtle exceptions due to band touching at the $\Gamma$ point~\cite{bergman2008band}.} are compact localized around individual hexagons~\cite{Mielke1992ka}, see Fig.~\ref{fig:1b}. 

We begin by studying the behavior of a single hole as we progressively depart from full polarization: \hs{1}{n}, where $n=1,2,3,\dots$ is the number of reversed spins. 
We observe that the reversed spins cluster around the hole, forming a polaron (akin to the behavior on the triangular lattice~\cite{zhang2018pairing}). 
The short distance physics of the {\kag} lattice is close to tree-like, and we find that the valence bond language of the same Hamiltonian on the Husimi cactus~\cite{kim2023exact} proves highly useful to model and understand the polarons in this work, as we demonstrate below. We therefore dub them RVB polarons to contrast them with the different, magnon-like description of polarons on the triangular lattice. The farther one departs from the polarized limit (i.e., the larger $n$), the less effective the valence bond language is, and the more the magnetic correlations become consistent with semiclassical {\rord} order. 

\begin{figure}
    \centering

    \centering
    \subfloat[]{%
        \raisebox{1.35em}{%
           \includegraphics[width=0.47\columnwidth]{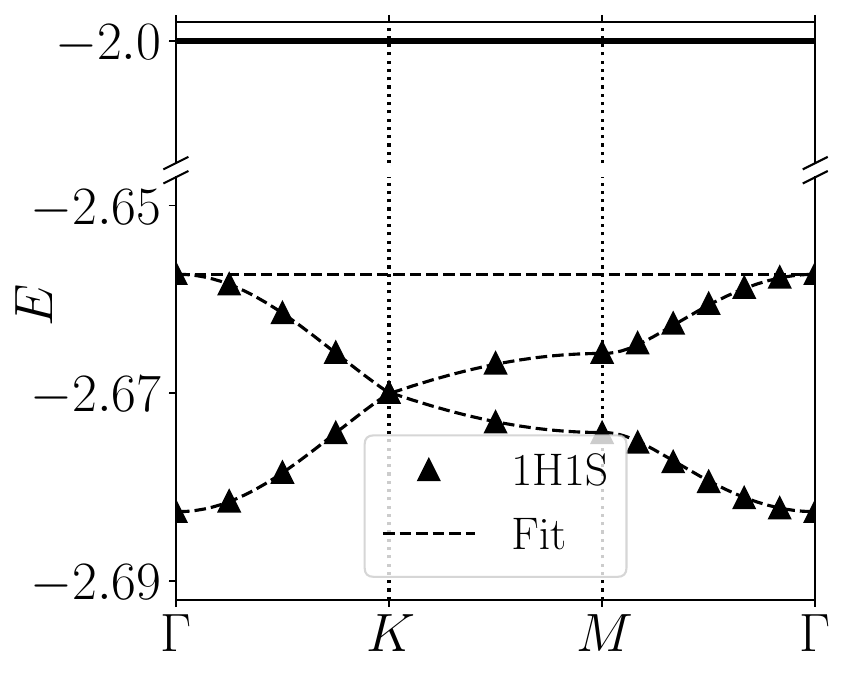}
           \label{fig:2a}
        }
    }
    \subfloat[]{%
    \raisebox{2em}
    {
        \includegraphics[width=0.22\linewidth]{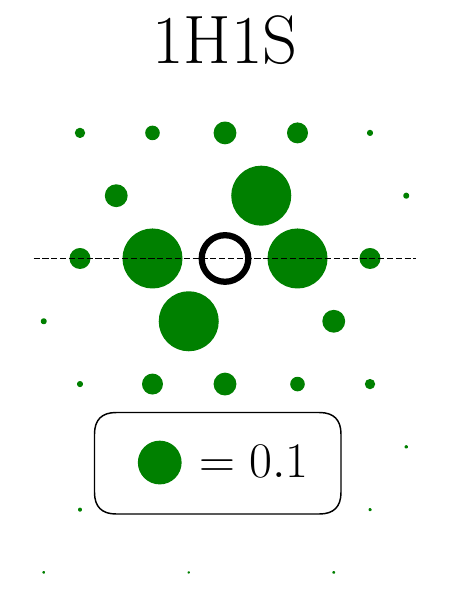}
        \label{fig:2b}
    }
    }
    \subfloat[]{%
    \raisebox{2em}
    {
        \includegraphics[width=0.22\linewidth]{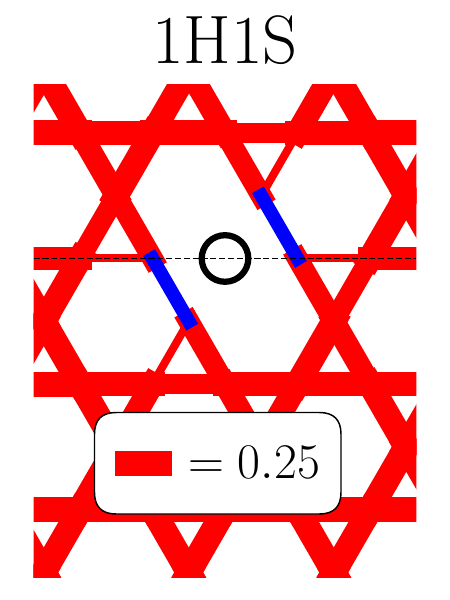}
        \label{fig:2c}
    }
    \hfill
    }
    
    \subfloat[]{%
        \raisebox{0.1em}{%
            \includegraphics[width=0.48\columnwidth]{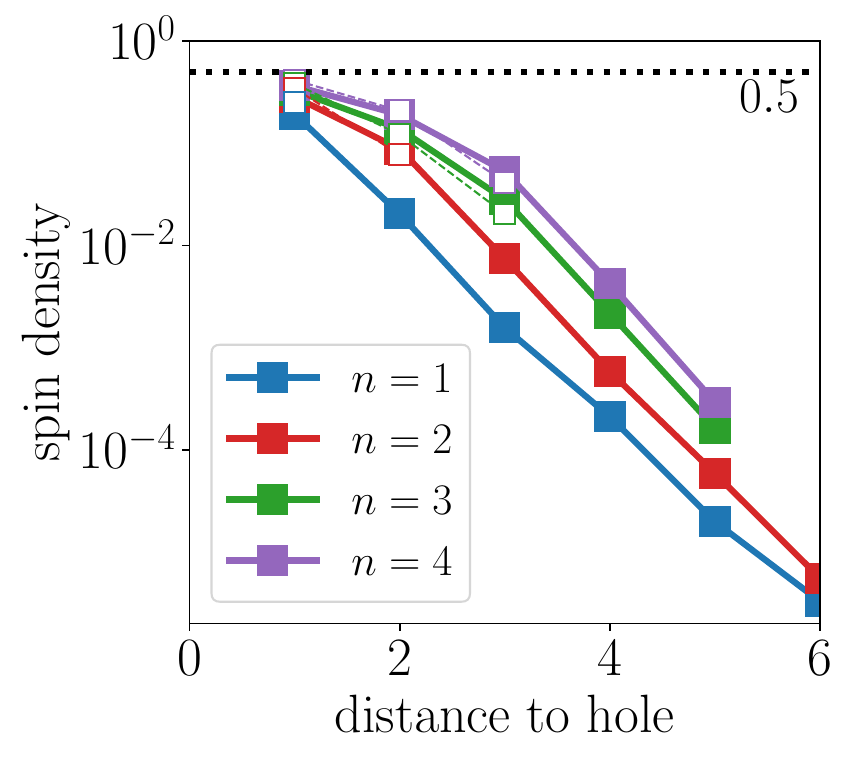}
            \label{fig:2d}
        }
    }
    \hfill
    \subfloat[]{%
        \includegraphics[width=0.46\columnwidth]{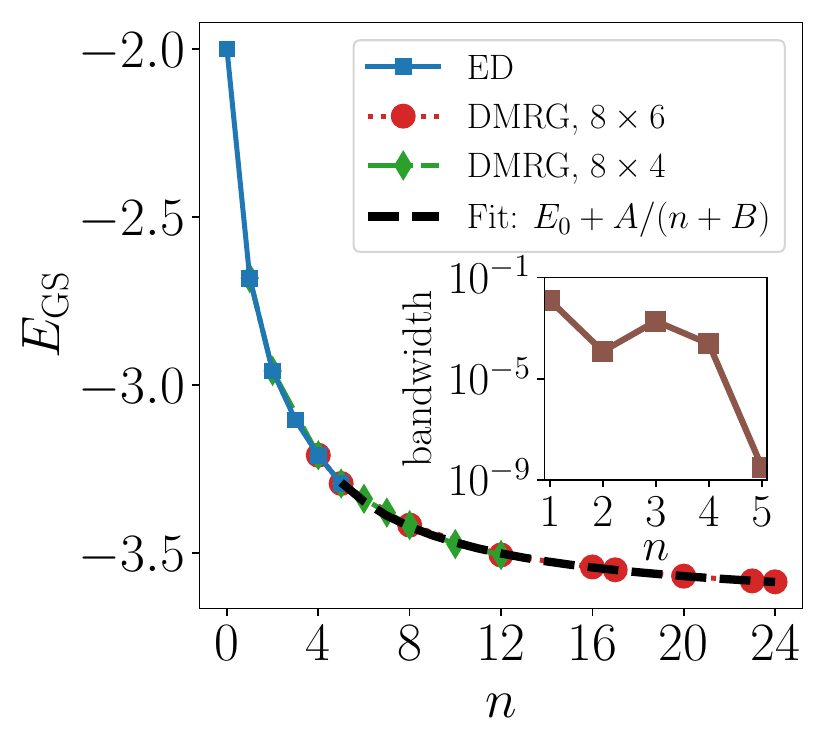}
        \label{fig:2e}
    }
    
    \caption{Band structure for the \hs{1}{1} system of size $12 \times 12$ (a), obtained using ED; the dashed line shows a single-particle tight-binding fit with $t_{\rm eff} \simeq -4.2\times10^{-3}$ and offset~$\simeq2.67$. Spin density (b) and spin correlations (c) (red ferro and blue antiferro, with width proportional to their strength), for fixed hole position. Spin density profile (d) along the horizontal line in panel (b), for \hs{1}{n} systems with $n=1,\ldots,4$, comparing ED (solid squares) with Husimi cactus cluster results (open squares). 
    The dependence of the GS energy of the \hs{1}{n} systems on $n$ is shown in panel (e), with the inset plotting our best estimate for the width of the lowest band from ED (see SI).} 
    \label{fig:2}
\end{figure}

Remarkably, for $n=1,2,3,4$ we find that the RVB polarons are delocalized -- contrary to the case of $n=0$ -- with an enhanced band mass (compared to the \hs{1}{1} polaron on the triangular lattice~\cite{zhang2018pairing}). ED shows a dispersive bottom band that is well-fit by an effective unfrustrated tight-binding model with hopping $t_{\rm eff} \simeq -4.2\times10^{-3}$, up to an overall energy offset (see Fig.~\ref{fig:2a} and SI~\cite{SI}). 
Note that only two out of the three bands of the \hs{1}{1} system are captured by the effective model; the third one is exactly flat at $E=-2$, split by a finite gap from the other two.

\begin{figure}
    \centering
    \includegraphics[width=\linewidth]{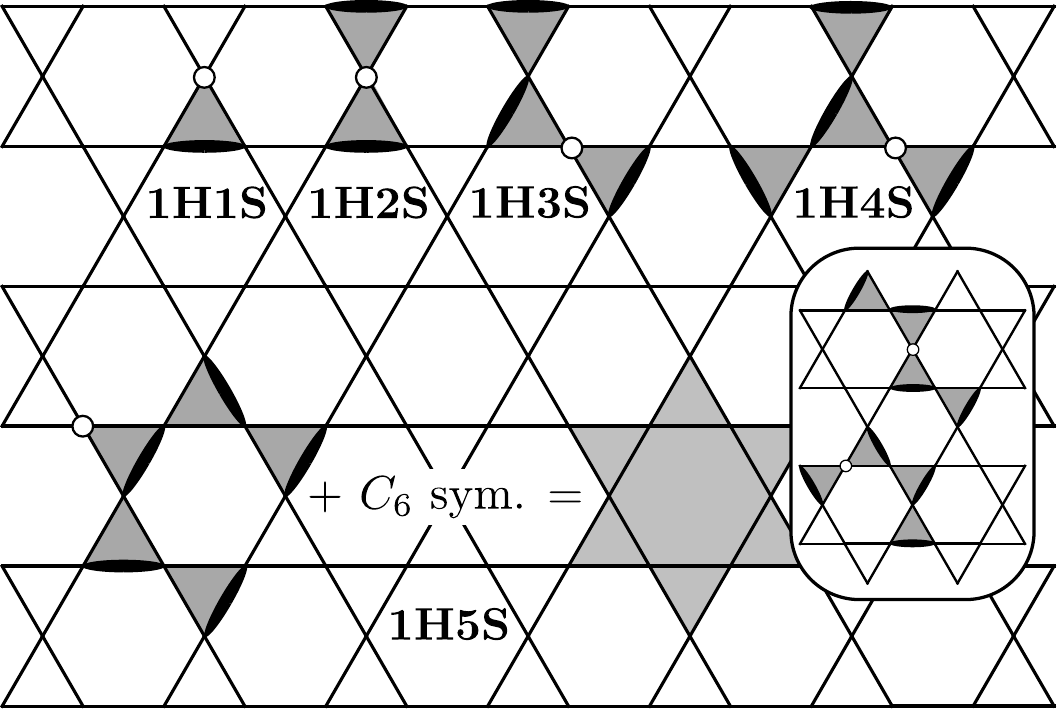}
    \caption{Examples of Husimi cactus clusters used in the main text to understand the nature of the polarons. All symmetry-inequivalent clusters are shown for \hs{1}{n}, with $1\le n \le 4$. For $n=4$, there are three possible configurations: a star-shaped one shown in the main figure, and two linear ones shown in the inset. For $n=5$, only the most relevant shape is shown; with its five symmetry-equivalent partners, this shape covers a $12$-site (shaded) region of the lattice. For each cluster, only one of the possible hole positions, and its corresponding valence bond arrangement, are shown. 
    }
    \label{fig:3}
\end{figure}

We look into the magnetic structure of the RVB polarons by computing the spin density and correlations after projecting the hole on a given site of the lattice (see Figs.~\ref{fig:2b}-\ref{fig:2d} and the SI). 
We are able to interpret these patterns in light of the valence bond behavior on the Husimi cactus~\cite{kim2023exact}: we consider all possible Husimi clusters that are compatible with the lattice and with the number of reversed spins (see Fig.~\ref{fig:3} for some examples, and the EM for technical details); we then compute the ground state of Hamiltonian~\eqref{eq:hamiltonian} on each of these clusters~\footnote{These states are superpositions of the hole at any of the sites on the cluster, and all electrons paired into nearest-neighbor singlets. A property of the Husimi cactus is that the singlet pattern is uniquely determined by the hole position.}; and we use them to create states for the full lattice by taking the tensor product with polarized spins on all other sites. 
For $n=1,2,3$ the clusters are unique up to lattice symmetries (see Fig.~\ref{fig:3}), and we find that the resulting states have excellent projection onto the exact GS of the full system ($88, 93, 90$\%, respectively). 
For $n=4$ there are three nonequivalent clusters (see Fig.~\ref{fig:3}), and we find that the star-shaped one has lowest cluster energy and largest projection onto the exact GS ($94$\%). 
These results show how the structure of the polarons on the {\kag} lattice is well-described by the local Husimi cactus valence bond physics -- hence the name RVB polarons. This is further confirmed by comparing the spin density (empty squares in Fig.~\ref{fig:2d}) and correlators (not shown) obtained from the Husimi cluster states described above and from the exact GS. 

These Husimi cluster states have finite hole support by construction. From Fig.~\ref{fig:2d} we see that the actual GS from ED exhibits exponentially decaying tails. The decay length appears to be asymptotically independent of the choice of cluster; it is indeed a property of the compact localized states that govern the physics away from the polaron (where the system is effectively fully polarized). 
The dispersive nature of the \hs{1}{n} lowest-energy bands (for $n=1,2,3,4$) tells us that the hole is able to move across the lattice whilst carrying such structured polaron along with it. The large band mass suggests that such cooperative motion involves high-order processes and correspondingly small amplitudes. 
%
%

\begin{figure}
    \centering

    \subfloat[]{%
        \includegraphics[width=0.47\columnwidth]{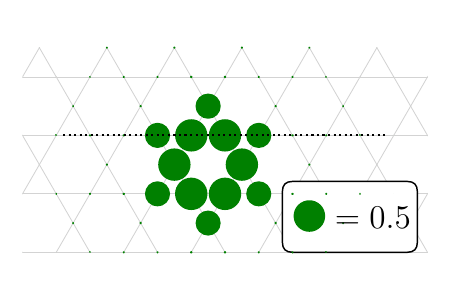}
        \label{fig:4a}
    }\hfill
    \subfloat[]{%
        \includegraphics[width=0.47\columnwidth]{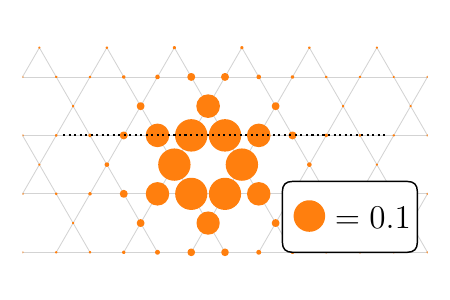}
        \label{fig:4b}
    }

    \subfloat[]{%
        \includegraphics[width=0.44\columnwidth]{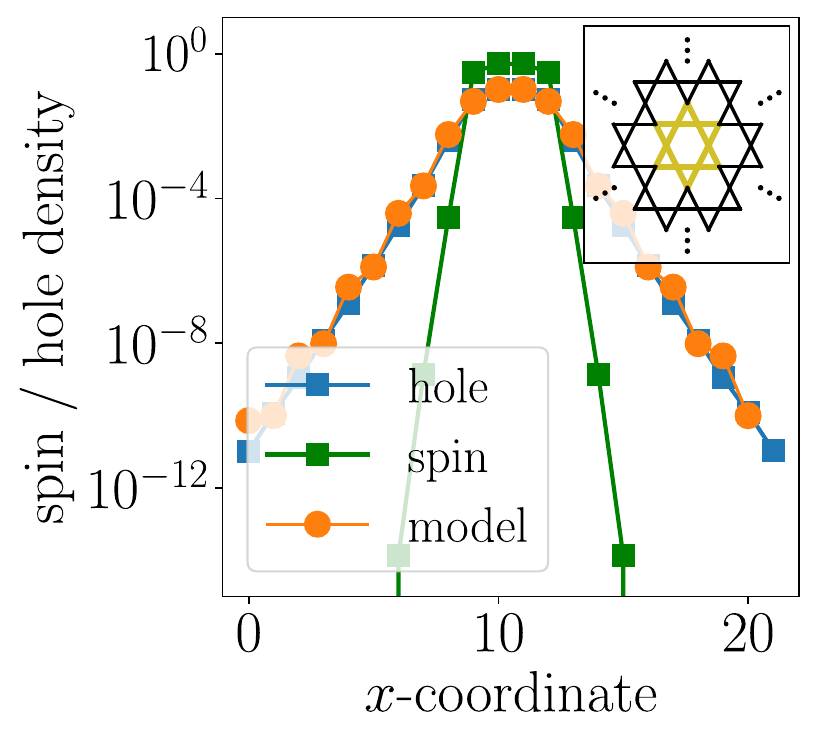}
        \label{fig:4c}
    }
    \hfill
    \subfloat[]{%
        \raisebox{0.1em}{%
            \includegraphics[width=0.5\columnwidth]{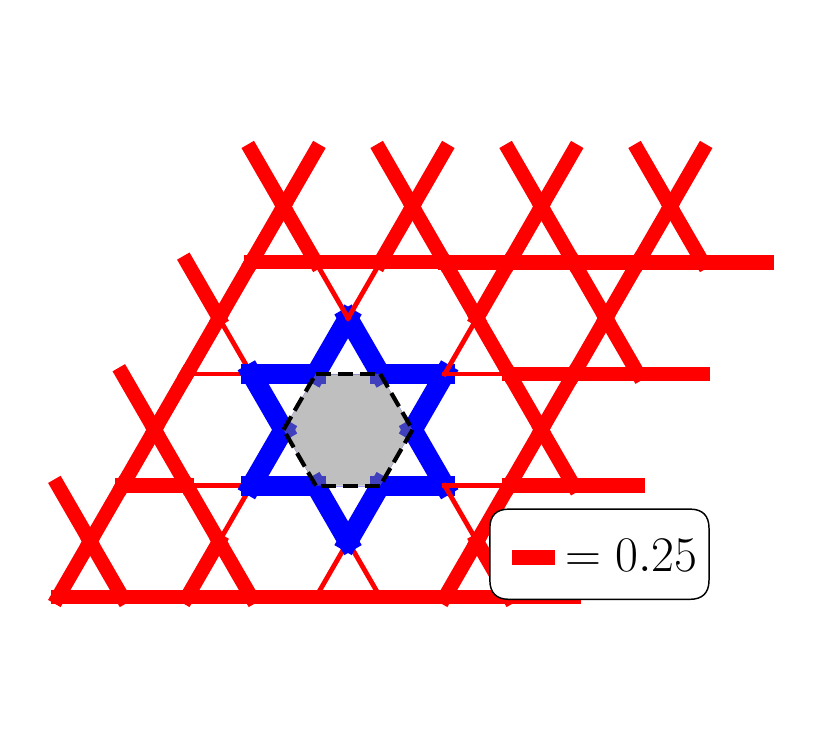}
        }
        \label{fig:4d}
    }\hfill

    \caption{Reversed spin (a) and hole (b) densities for one of the seemingly-localized \hs{1}{5} GSs. 
    Profile cuts are shown in panel (c). The hole density is well-described by a simple model of a spinless fermion (see inset and main text) with unfrustrated hopping on a $12$-site cluster ($t=-1$, yellow edges), embedded on a periodic {\kag} lattice with frustrated hopping ($t=1$, black edges). 
    The spin correlators (red for ferro, blue for antiferro) are shown in panel (d), with the thickness of the lines denoting their strength. Density results were obtained using DMRG and spin correlators using ED (see EM). A vanishing pinning potential was used to single out a given location for the RVB polaron in the GS.}
    \label{fig:4}
\end{figure}

\emph{Strong self-trapping} --- 
The behavior of the system changes dramatically for $n\geq5$. The polaron bandwidth suddenly drops by at least $6$ orders of magnitude ($< 10^{-8}$ in a $4\times 4$ system of $48$ sites, with $16$ near-degenerate GSs); to the best of our numerics it is indistinguishable from a flat band of localized states. 
The reversed-spin density appears to be supported by a $12$-site cluster symmetrically arranged around a hexagonal plaquette (Fig.~\ref{fig:4a}), while the hole density appears to extend farther (Fig.~\ref{fig:4b}). Both exhibit a pronounced exponential decay beyond the $12$ sites (Fig.~\ref{fig:4c}), as confirmed by DMRG and ED~\footnote{We added a small uniform pinning potential around a given hexagonal plaquette to select one of the GSs (see EM).}. 
We find a ratio of approximately $4$ between the spin and hole density correlation lengths, which can be qualitatively understood from the fact that reversed spin motion is mediated by hole motion, and it takes an excursion with at least $4$ hole hops starting from the $12$-site cluster to move a reversed spin out of it. 
The spin correlators are shown in Fig.~\ref{fig:4d}. 

To gain insight into the sudden change of behavior of the \hs{1}{5} case, we consider again the effective modeling in terms of Husimi clusters employed earlier for $n<5$. Similarly to $n=4$, multiple cluster shapes are possible, and their energy is lowest the more ramified their structure. However, for $n=5$ the most ramified structure does not provide the largest overlap with the exact GS. This is achieved instead by a maximally curved linear arrangement of triangles, illustrated in Fig.~\ref{fig:3}. 
This apparent contradiction is explained by the fact that there are six symmetry-related such clusters around any hexagonal plaquette of the lattice, and the corresponding six Husimi cactus states are not orthogonal. By restricting the Hilbert space of the system to the subspace spanned by these six states, and then obtaining the GS of the Hamiltonian within it (see EM), we find indeed that the GS energy is lower than that of all other Husimi clusters, and the GS wavefunction takes the form of a symmetric superposition of all six states, spreading across all $12$ sites around the hexagonal plaquette (see SI). Its overlap with the exact GS is once again excellent ($90\%$).

By resonating around a hexagonal plaquette, it seems that the \hs{1}{5} RVB polaron acquires a possibly-infinite mass, undergoing localization reminiscent to the compact-localized hexagonal states of the spinless single-particle problem~\cite{Mielke1992ka}. 
Indeed, our results suggest that the reversed spins form a background that allows the hole to delocalize across the $12$ sites, whereas its density is heavily suppressed outside of the region. We can then assemble an ad hoc tight binding model for a spinless fermion on the {\kag} lattice, with frustrated hopping ($t=1$) everywhere except for bonds between the $12$ sites, where we set unfrustrated hopping ($t=-1$), see inset of Fig.~\ref{fig:4c}. The ground state wavefunction of this simple toy model has particle density in excellent agreement with the hole density of the original problem, as illustrated by a comparison of density profile cuts in Fig.~\ref{fig:4c}. 
%
%

\emph{Unpolarized state and $\sqrt{3} \times \sqrt{3}$ correlations} --- 
The study of \hs{1}{n} systems with $n>5$ is beyond the reach of our ED. DMRG results suggest that their behavior is similar to the case $n=5$ discussed above (see SI). 
Once again we find evidence of a compact support region for the reversed spins, with the hole density also exponentially localized to it -- albeit less tightly than the reversed spins~\footnote{This result is further supported by ED on a finite region of the lattice given by the sites where the reversed spins are localized. The correlators obtained from the ED GS are indeed in good agreement with the ones obtained from DMRG on the full system (see SI).}. 

The GS energy is plotted in Fig.~\ref{fig:2e}, and shows scaling $\simeq E_0 + A/(n+B)$ at large $n$ (where $E_0 = -3.68 \pm 0.01$, $A = 2.26 \pm 0.13$ and $B=0.84 \pm 0.23$, from a fit performed over $5 \leq n \leq 24$). This is consistent with the reversed spins forming an effective quantum well confining the hole to a region of linear size $\propto \sqrt{n}$ ($n$ reversed spins live in a region formed by $n$ triangles on the {\kag} lattice). 

As the size of the reversed spin region grows, it encompasses more hexagonal loops of the lattice and the effective modeling in terms of Husimi clusters becomes less and less useful. We see that this `frustration on frustration' leads to the appearance of magnetic correlations consistent with semiclassical {\rord} order as we approach the unpolarized limit. 
To show this, we evaluate the structure factor
\begin{equation}\label{eq:S_k}
    S(\mathbf{k}) = \frac{1}{N^2} \sum_{i, j} \langle \mathbf{S}_i \cdot \mathbf{S}_j \rangle \mathrm{e}^{\mathrm{i} \mathbf{k} \cdot (\mathbf{r}_i - \mathbf{r}_j)}
\end{equation}
for the GS obtained either by ED (on a $3\times 3$ system) or by DMRG (on cylinders with $L_y=3$ and $L_x$ integer multiple of $3$). In all cases, we find 
peaks at the $K$ points, suggestive of {\rord} order (see SI for intensity plots of $S(\mathbf{k})$ in the entire BZ)~\footnote{Linear system sizes multiple of $3$ were chosen to ensure commensurability.}. 

\begin{figure}
    \centering
    \includegraphics[width=0.75\columnwidth]{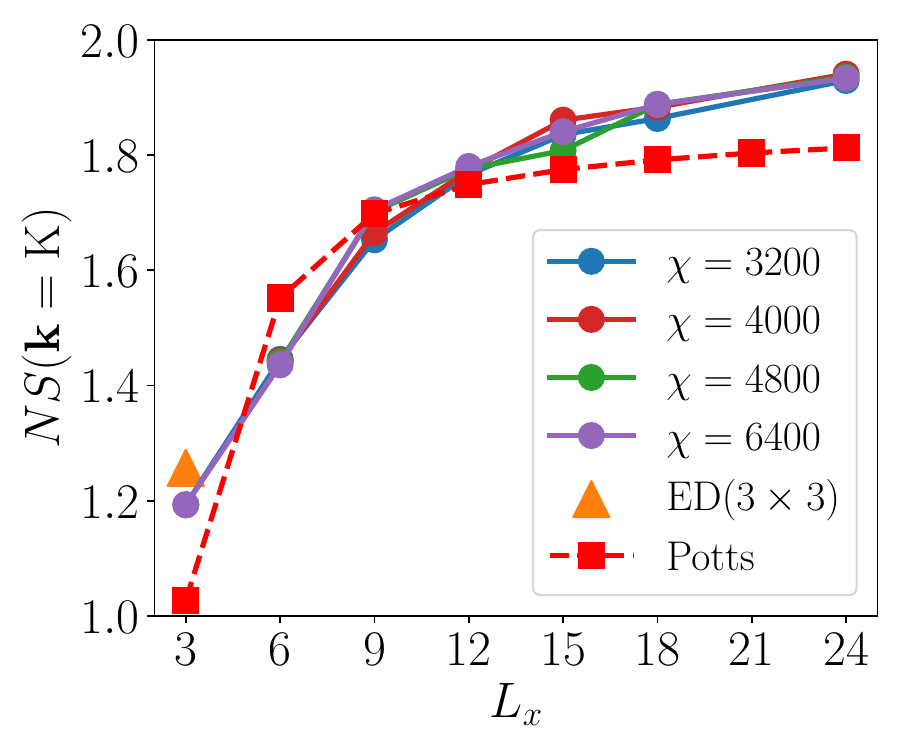}
    \caption{Rescaled peak intensity at the $K$ point for our unpolarized system on lattices of size $L_x \times 3$, with $6\le L_x \le 24$, obtained using DMRG (circles joined by solid lines). We also show the $3 \times 3$ result obtained using ED (yellow triangle).  
    We compare these results with the behavior of the $3$-state Potts model on lattices of size $L_x \times 3$, with $6\le L_x \le 24$ (squares and dotted lines), which in $d=2$ is known to sit on a KT transition between algebraic and long-range magnetic correlations. 
    }
    \label{fig:5}
\end{figure}

We plot the intensities of the peaks multiplied by system size, $NS(\mathbf{k}=K)$ in Fig.~\ref{fig:5}, showing that $S(\mathbf{k}=K)$ decreases more slowly than $1/N$. For comparison, we present results from classical simulations of the $3$-state Potts model on systems of equivalent size. 
This model is useful~\cite{huse1992classical,chern2013dipolar} in that it sits at the KT transition separating a {\rord} ordered and an algebraically correlated phase, and it can thus provide a natural benchmark for an analysis of semiclassical correlations. It reflects the low-temperature correlations of a classical XY kagome magnet exactly, and -- appropriately perturbed -- those of the classical Heisenberg magnet approximately. Its straightforward numerical tractability allows the analysis of systems of the finite sizes to which our other numerical methods are restricted. 
The Potts model exhibits similar values of $S(\mathbf{k}=K)$ when the classical correlators are scaled as appropriate for semiclassical behavior (i.e, by $(S^{z})^2=1/4$ instead of $S(S+1) = 3/4$, see EM) and a similar dependence of $S(\mathbf{k}=K)$ on system size.  
Spin correlations in our model exhibit slow decay and oscillations also consistent with {\rord} order (see SI). Indeed, the slightly larger value, and faster growth, of the correlations with system size in Fig.~\ref{fig:5} for our problem compared to the Potts model hints at the possibility of a single hole inducing long-range semiclassical {\rord}  magnetic order. 
We note however that the identification of ordered states on the kagome lattice is notoriously challenging and we cannot rule out other possibilities, in particular if they involve large unit cells or more complex order parameters. 
%
%

\emph{Conclusions} --- 
We presented a study of kinetic frustration in a singly-doped Mott insulator on the {\kag} lattice. 
Close to full polarization, we uncover a complex interplay between flat-band localization and polaron formation resulting in non-monotonic band-mass behavior. We developed a description of the polarons in terms of resonant-valence-bond states characteristic of the same system on a Husimi cactus~\cite{kim2023exact}. 
When these RVB polarons are large enough to encircle hexagonal plaquettes on the lattice, we found that their band mass diverges within our numerical precision, and they appear to re-localize in a disorder-free system. 
As we moved towards the unpolarized limit, the RVB polaron description becomes less useful and the spin correlations appear to morph toward semiclassical {\rord} order. 

Valence bond physics has been shown to play a key role in kinetic frustration on line-graph lattices~\cite{Mielke1992,Brandt1992} such as the 2D checkerboard and 3D pyrochlore lattices. In these cases, the unpolarized limit of our problem results in an RVB spin liquid state~\cite{Tasaki1993,glittum2025resonant}. Interestingly, the fully polarized limit of these lattices hosts once again flat bands and compact localized states. Therefore it will be interesting to extend our study of the behavior of RVB polarons and their non-monotonic band mass in these systems as one varies polarization. 

Another important direction for future work is the case when multiple holes are present in the system. This opens the door to polaron-mediated interactions between the holes, and possible binding, pairing and phase separation effects. Our preliminary study of the \hs{2}{10} system indeed suggests an effective attraction between the holes. 

One also wonders whether the addition of farther-range hopping terms could lead to itinerant bipolarons, as recently found on the triangular lattice~\cite{zhang2018pairing}. 


\textbf{Data Availability Statement} --- 
The data that support the findings of this study are available from the corresponding author upon reasonable request. 
%
%

\textbf{Acknowledgements} --- 
We thank C.~D.~Batista, C.~Glittum, Jos\'{e} J. Baldov\'{i}, F.~Pollmann, and A.~Wietek for insightful discussions. We are particularly indebted to C.~Glittum for sharing with us preliminary results on an $L_x=3 , \, L_y=3$ kagome system. 
This work was supported in part by the Engineering and Physical Sciences Research Council (EPSRC) grant No.~EP/V062654/1 (CC), and the Deutsche Forschungsgemeinschaft via the cluster of excellence ctd.qmat (EXC 2147, project-id 390858490) and  SFB 1143 (Project-ID No. 247310070).
%
%

\section*{End Matter}

\subsection*{Methods}
We used exact diagonalization (ED) to obtain our main results for \hs{1}{n}, $n=1,\ldots,5$. We were able to access sizes up to $12\times12$ (namely, $3L_xL_y=432$ sites) for $n=1$, up to $9\times9$ for $n=2$, up to $6\times6$ for $n=3$, up to $5\times5$ for $n=4$, and up to $5\times4$ for $n=5$, all with periodic boundary conditions. As the $K$ points in the BZ can only be accessed for linear system sizes that are integer multiple of $3$, it was not possible to read off directly the band maxima in our largest systems for $n=4,5$. Rather than presenting the $3\times3$ or $6\times3$ data, where finite size effects are severe, we chose to present the $5\times5$ and $5\times4$ data, respectively for $n=4$ and $n=5$, as a lower bound estimate of the bandwidth (see SI). 

To obtain the data shown in Fig.~\ref{fig:4d}, we added a pinning potential $\epsilon\sum_{i\in p}\hat{n}_i$ with $p$ identifying the $6$ sites around a given hexagonal plaquette, and $\epsilon=10^{-6}$. This allowed us to select one of the $12$-site \hs{1}{5} GSs in our ED. We checked that the chosen value of $\epsilon$ was sufficiently small not to affect the nature of the selected GS. We further verified this against density matrix renormalization group (DMRG) results (see SI). 

For $n\ge5$, we conducted DMRG calculations in YC-geometries with system sizes $8\times 4$ and $8\times6$ using the TenPy package~\cite{tenpy2024}, apart from $n=5$ where an additional $12\times4$ simulation was performed (Fig.~\ref{fig:4c}). In all DMRG calculations, we used cylindrical geometry with open boundary conditions in the $x$ direction. DMRG calculations were considered converged when the energy difference between two consecutive sweeps fell below $10^{-8}$, and the entanglement entropy difference below $10^{-6}$, with a maximum sweeping count of $200$. The bond dimension was set to $\chi=6400$. We contrasted results on different system sizes to check that they agree within statistical error, up until the point where the polaron is large enough for boundary effects to become important. 

In the unpolarized sector, we used ED for $L_x=L_y=3$ (with periodic boundary condition), and DMRG for $L_y=3$ and $L_x$ ranging from $6$ to $24$ (with open boundary conditions in the $x$ direction); {\rord} order requires $L_x$ and $L_y$ to be integer multiple of $3$ to avoid incommensurability issues. 
The calculations were considered converged when the energy difference between two consecutive sweeps fell below $10^{-7}$, and the entanglement entropy difference below $10^{-5}$, with a maximum sweeping count of $200$.
%
%

\subsection*{Husimi cluster states}
In order to gain insight into the structure of the \hs{1}{n} polarons, we identify the possible Husimi cactus shapes consisting of $n$ adjacent triangles on the {\kag} lattice (see Fig.~\ref{fig:3} in the main text). 
We diagonalize the Hamiltonian in Eq.~\eqref{eq:hamiltonian} restricted to these clusters, and identify the GSs. In agreement with Ref.~\onlinecite{kim2023exact}, the GS on each of these clusters is a unique superposition (with real positive amplitudes) of all possible hole positions across the cluster and the electrons paired into nearest-neighbor singlets. These GSs are then extended to states of the full system ($|\psi_{\mathrm{HC}}\rangle$) by taking their tensor product with polarized spins on all other sites. 
Crucially, for $n=1, 2, 3, 4$, the states $|\psi_{\mathrm{HC}}\rangle$ are mutually orthogonal. 
The projection values (absolute values of the inner products) onto the GS of the original system (obtained numerically using ED) are reported in the main text, after accounting for translation and rotation invariance. 
For $n=4$ there are three symmetry-inequivalent Husimi cluster shapes, as illustrated in Fig.~\ref{fig:3}. In a $5\times5$ system, we find that the one lowest in energy is the most ramified (tree-like) one, and it leads to a $93\%$ projection onto the GS, as reported in the main text. The other two shapes have higher energy and lead to $9\%$ and $0.4\%$ projection (top and bottom shape in the inset of Fig.~\ref{fig:3}, respectively). 

For $n=5$, there are many different Husimi cluster shapes. The maximally curved linear arrangements discussed in the main text are the only ones where the Husimi cactus states $|\psi_\mathrm{HC}^{(i)}\rangle$, $i=1,\ldots,6$ -- corresponding to different clusters obtained by rotations around the central hexagon -- are non-orthogonal (the dimer-singlet basis is over-complete on the $12$-site cluster). While each such state has energy $-3.03$, their superposition $|\psi_\mathrm{12-sites}\rangle =\sum_{i=1}^6(-1)^i|\psi_\mathrm{HC}^{(i)}\rangle$ in fact has lower energy: $\langle \psi_\mathrm{12-sites}|\mathcal{H}|\psi_\mathrm{12-sites}\rangle=-3.10$. This is lower than the energy of the most ramified configuration, $E=-3.08$. We therefore take $|\psi_\mathrm{12-sites}\rangle$ as our Husimi cluster variational \hs{1}{5} ground state, and find that it has an excellent overlap with the exact result from ED, as reported in the main text. By comparison, the most ramified configuration has only a $0.1\%$ projection (in a $4\times4$ system). A further discussion of this state can be found in the SI, where an in-depth study of the $12$-site cluster is also presented. 
%
%

\subsection*{Potts Model}
The data presented in Fig.~\ref{fig:5} include our study of the $3$-state Potts model, with Hamiltonian $H_\mathrm{Potts}=\sum_{\langle ij\rangle}\delta_{\sigma_i\sigma_j}$, $\sigma_i=0, 1, 2$. The ground state ensemble is extensively degenerate and consists of all states without identical nearest neighbors. 
We used Monte-Carlo simulations with loop updates to sample the ground states at zero temperature. The algorithm starts with a configuration satisfying the nearest-neighbor constraint and iteratively updates it by identifying closed loops with alternating states and exchanging them. We present results with open boundary conditions in the $x$-direction in Fig.~\ref{fig:5}. In calculating the spin-spin correlations and the spin structure factor in Eq.~\eqref{eq:S_k}, we treat each of the Potts states as classical spins of length $1/2$ (as appropriate for semiclassical order). Therefore, identical Potts states give $\mathbf{S}\cdot\mathbf{S}=1/4$ whereas different states give $\mathbf{S}\cdot\mathbf{S}=-1/8$. This choice guarantees that a ferromagnetic classical and a ferromagnetic quantum spin-$1/2$ system give the same value of $S(\mathbf{0})$. 

The Potts model is known to sit at the KT transition separating a {\rord} ordered phase and an algebraically correlated phase~\cite{huse1992classical,chern2013dipolar}, and it provides a natural benchmark for an analysis of semiclassical {\rord} correlations. 
In the SI, we further show spin correlation function data suggesting that kinetic frustration on the {\kag} lattice leads to more pronounced {\rord} order than the Potts model (at least on our quasi-1D cylindrical geometry). 
%
%

\bibliographystyle{apsrev4-2}
\bibliography{ref}

\clearpage
\onecolumngrid

\setcounter{section}{0}
\setcounter{figure}{0}
\setcounter{table}{0}
\setcounter{equation}{0}

\renewcommand{\thesection}{S\arabic{section}}
\renewcommand{\thefigure}{S\arabic{figure}}
\renewcommand{\thetable}{S\arabic{table}}
\renewcommand{\theequation}{S\arabic{equation}}

\begin{center}

{\large \textbf{Supplemental Material for ``Kinetic kagome magnetism: from self-trapping RVB polarons to semiclassical correlations''}}

\end{center}
%
%

\section{Additional exact diagonalisation results}
\begin{figure}[bp]
    \centering
    \includegraphics[width=0.16\linewidth]{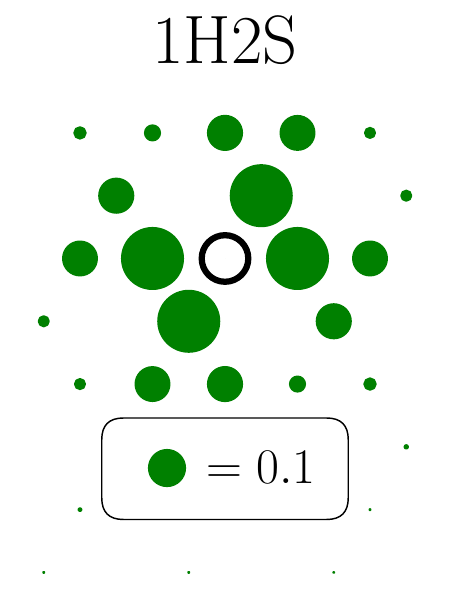}
    \includegraphics[width=0.16\linewidth]{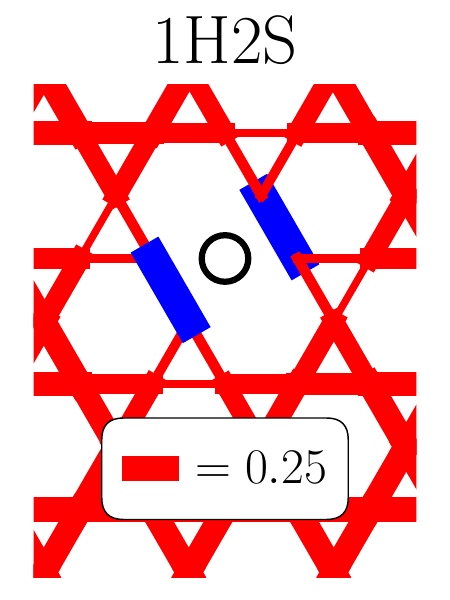}
    \includegraphics[width=0.16\linewidth]{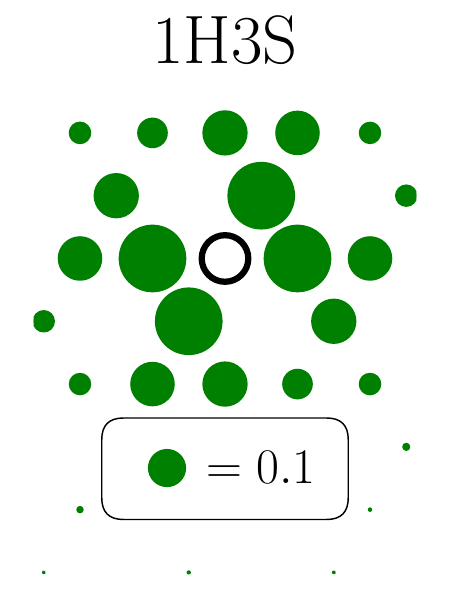}
    \includegraphics[width=0.16\linewidth]{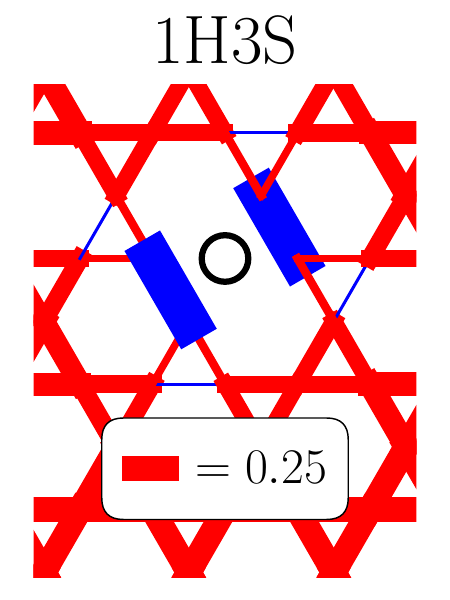}
    \includegraphics[width=0.16\linewidth]{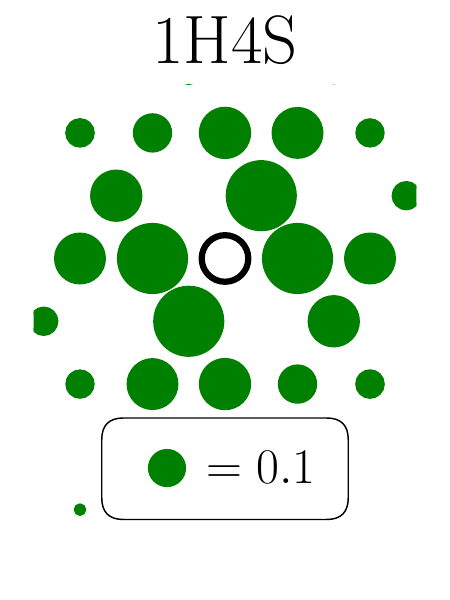}
    \includegraphics[width=0.16\linewidth]{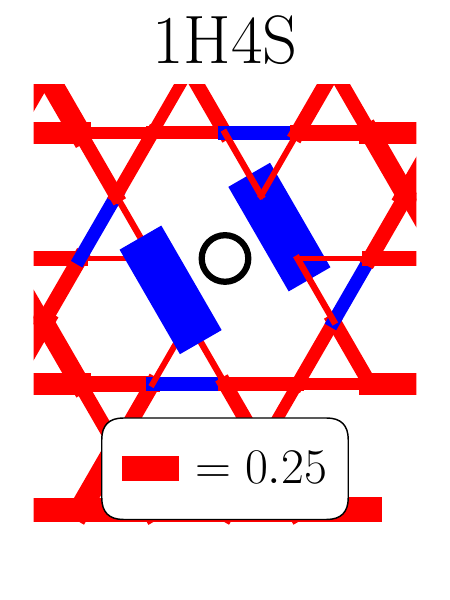}
    \caption{Reversed-spin density and nearest-neighbor spin-spin correlator in \hs{1}{n} systems with $n=2, 3, 4$ after projecting the hole onto a given site, obtained via ED for $9\times9$, $6\times6$ and $4\times4$ systems, respectively. Red (blue) bonds indicate positive (negative) correlations, and the thickness of the lines denotes their strength.}
    \label{fig:S1}
\end{figure}

\begin{figure}[t]
    \centering
    \includegraphics[width=0.4\linewidth]{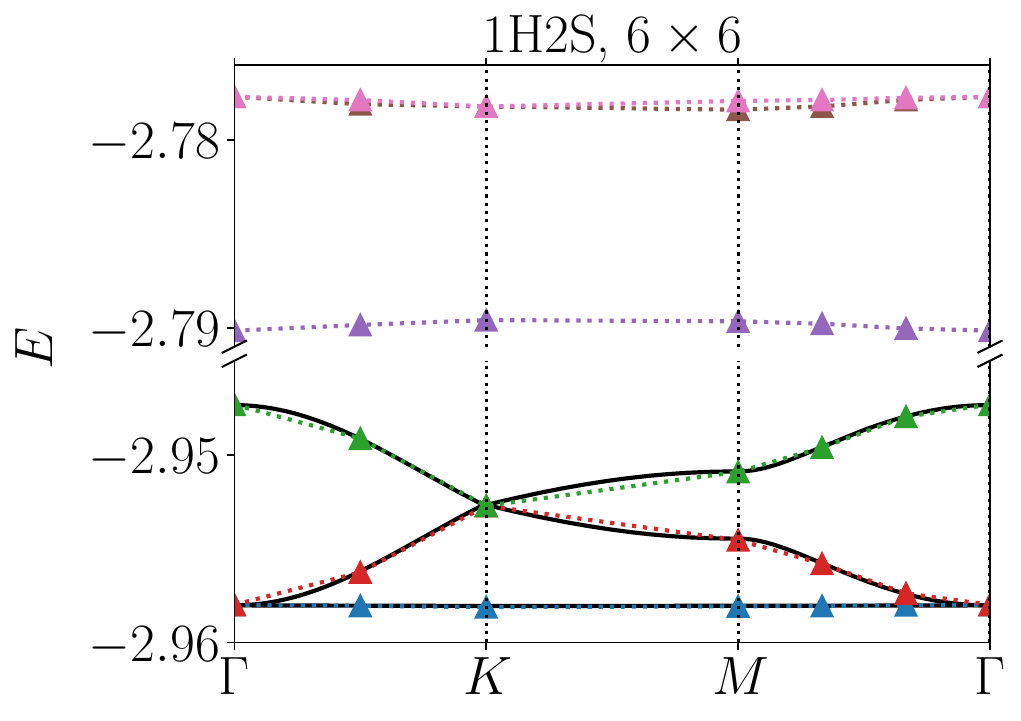}
    \includegraphics[width=0.4\linewidth]{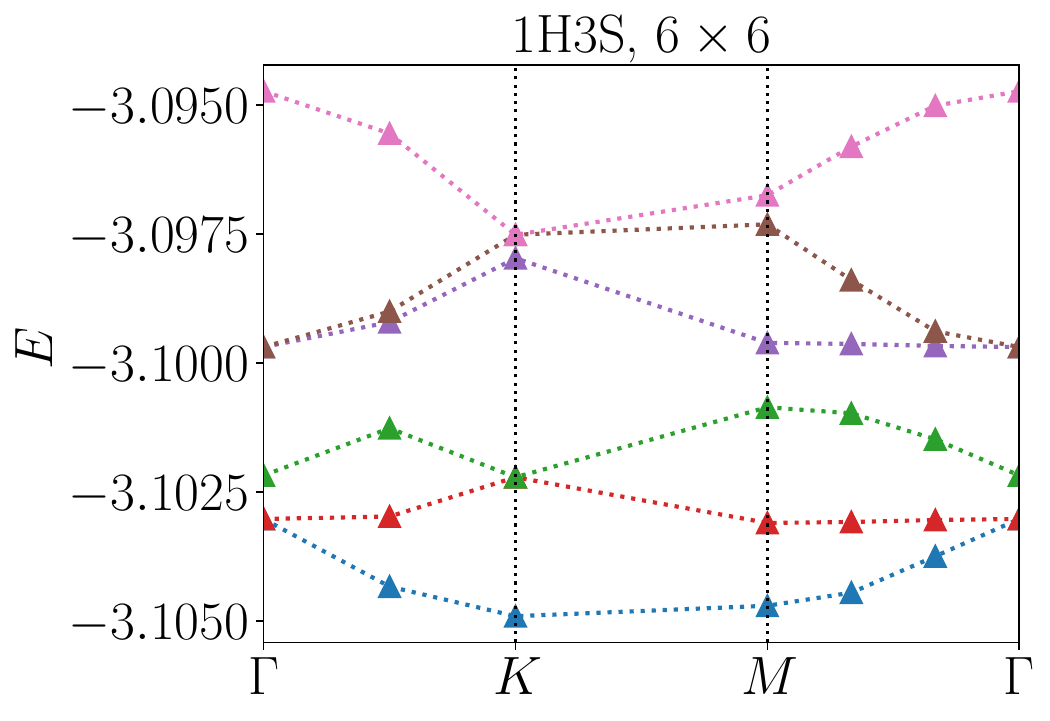}

    \includegraphics[width=0.4\linewidth]{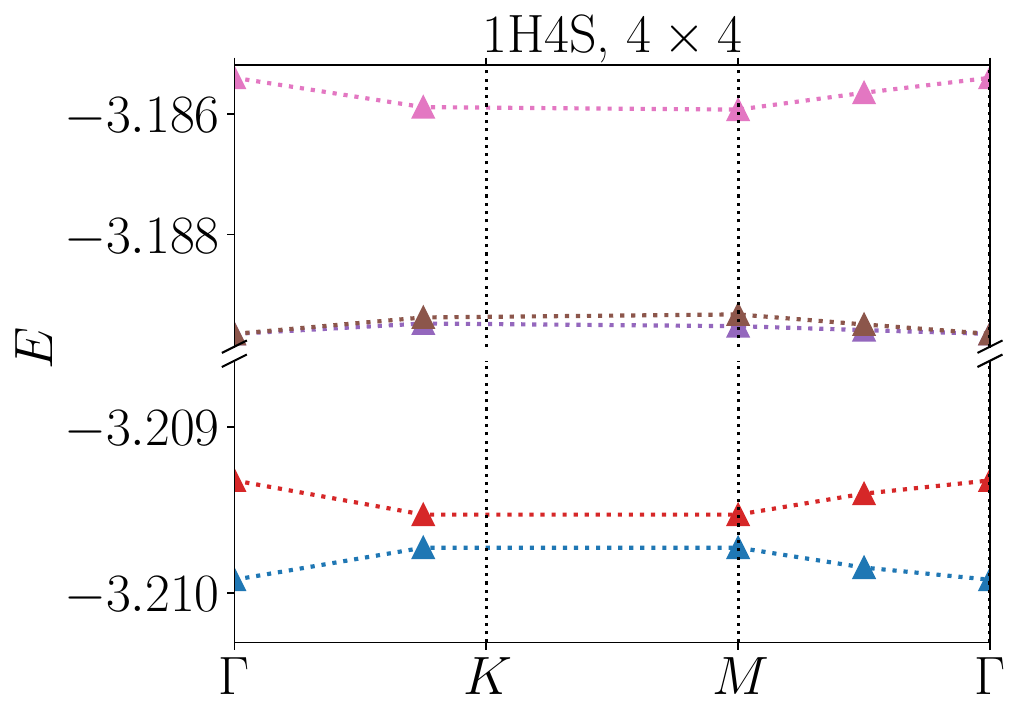}
    \includegraphics[width=0.4\linewidth]{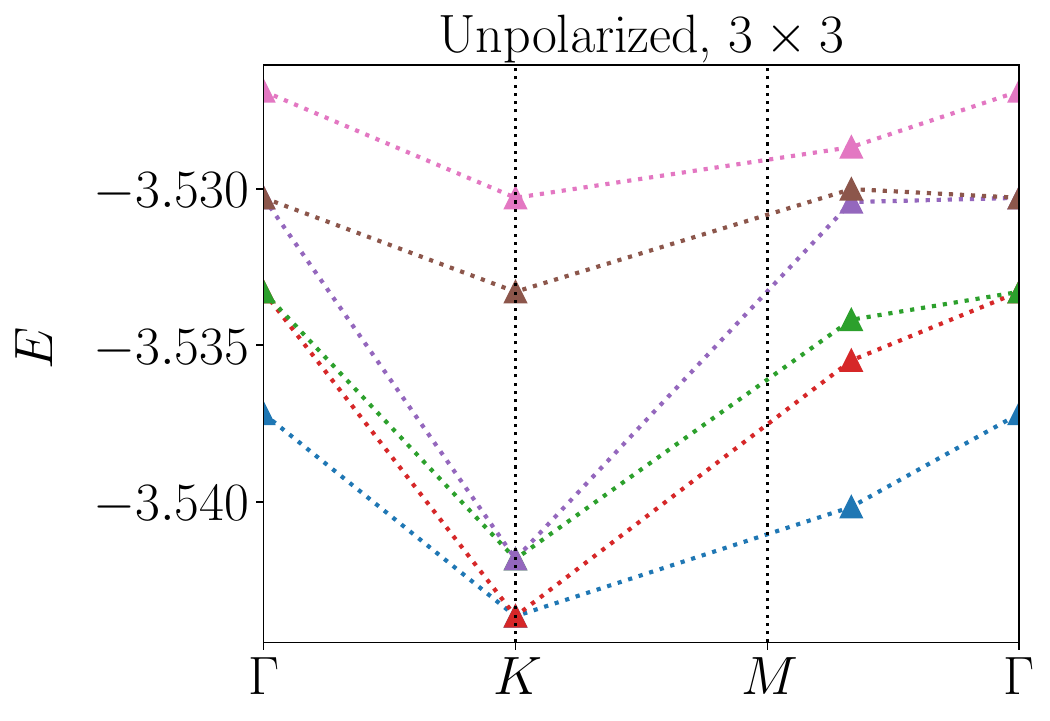}

    \caption{Lowest $6$ bands for different systems: \hs{1}{n} with $n=2, 3, 4$ as well as the unpolarized sector (with different system sizes, as indicated above each panel). These plots are made along the paths in reciprocal space shown in Fig.~\ref{fig:SBZ} below. For \hs{1}{2}, we also show fits with a single-particle tight-binding model with nearest- and next-nearest-neighbor hopping (black solid lines).}
    \label{fig:Sband}

\end{figure}

\begin{figure}

    \centering

    \subfloat[]{
        \centering
        \includegraphics[width=0.244\linewidth]{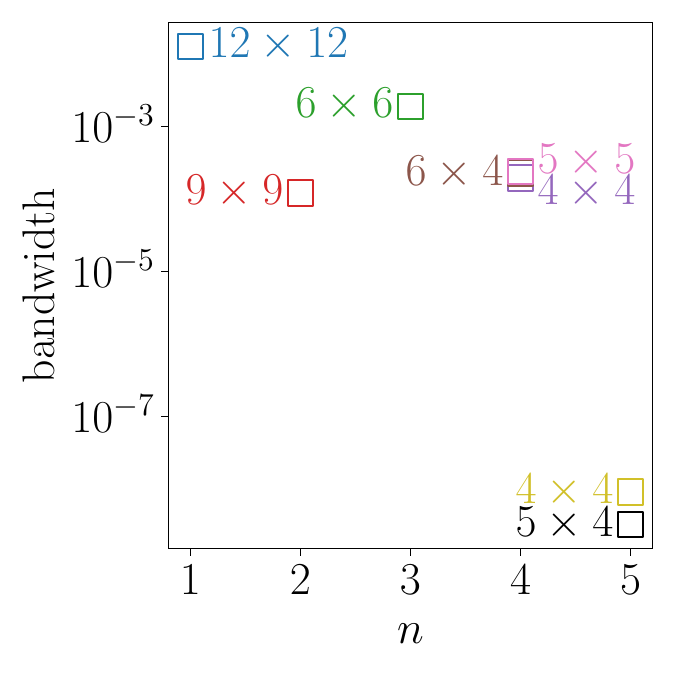}
        \label{fig:Sbw1}
    }
    \subfloat[]{
        \centering
        \includegraphics[width=0.41\linewidth]{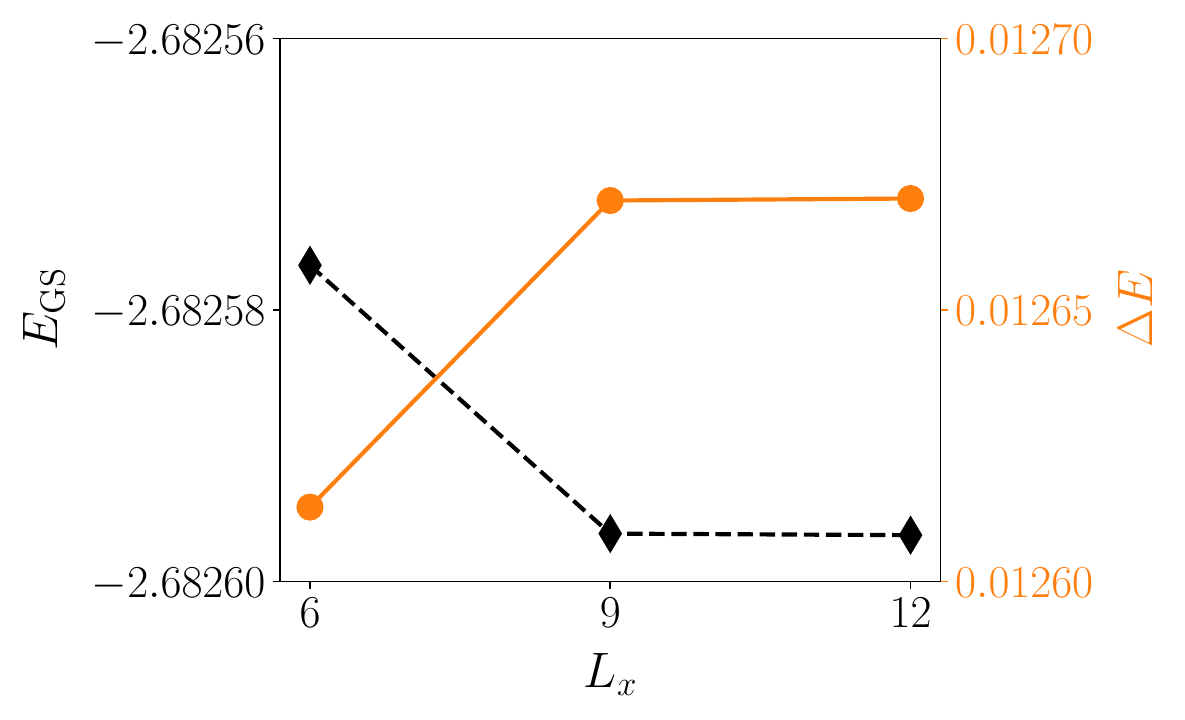}
        \label{fig:Sf1}
    }
    \subfloat[]{
        \centering
        \raisebox{0.2em}{
        \includegraphics[width=0.266\linewidth]{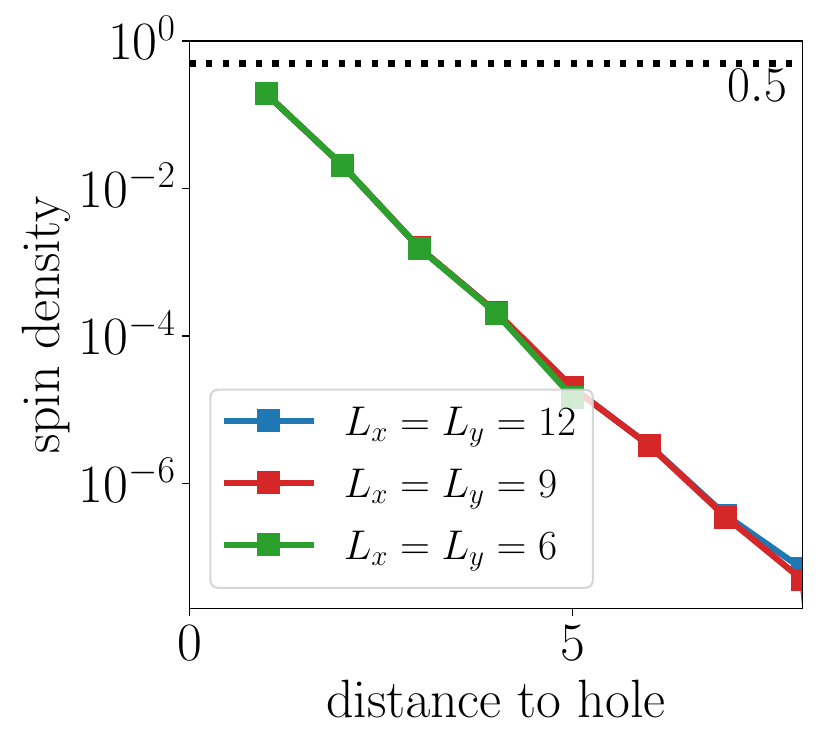}
        }
        \label{fig:Sf2}
    }
    
    \caption{All bandwidth data computed using ED for \hs{1}{n} systems with $L_x>3, L_y>3$ plotted together (a), expanding on the inset of Fig.~\ref{fig:2e} in the main text (where we only show the bandwidth data for each value of $n$ corresponding to the largest system size we are able to access). For $n=4,5$ different system sizes were used, as indicated in the panel. Finite size dependence of the \hs{1}{1} system: (b) ground state energy $E_\mathrm{GS}$ and bandwidth $\Delta E$ for $L_x=L_y=6, 9, 12$; (c) hole-spin correlations.} 
    \label{fig:Sfs}

\end{figure}

\begin{figure}
    \centering
    \includegraphics[width=0.32\linewidth]{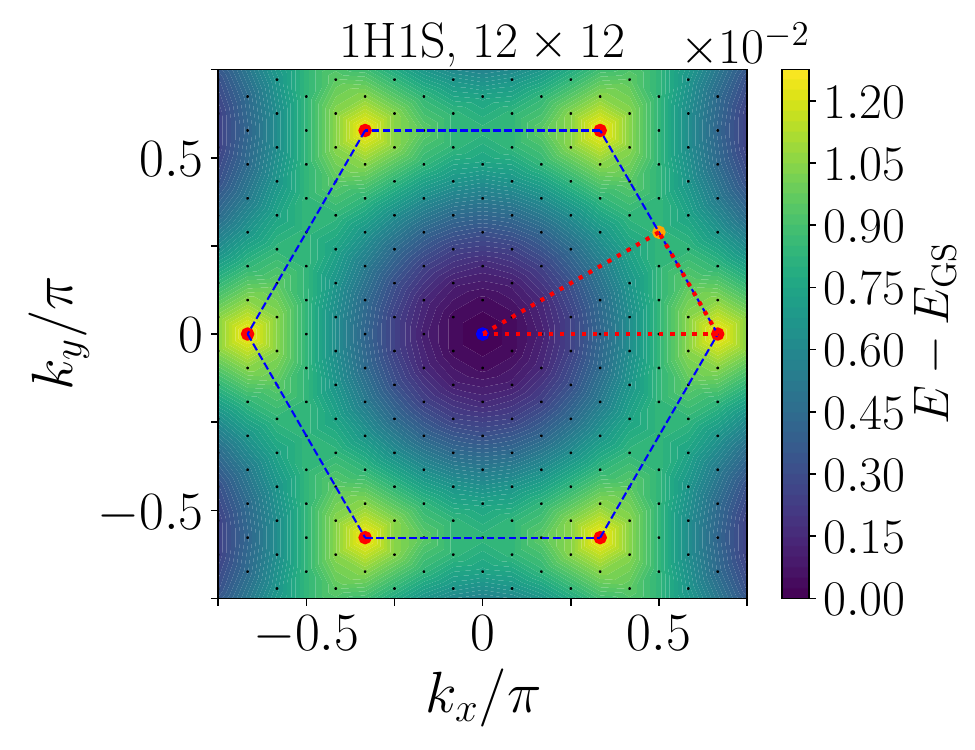}
    \includegraphics[width=0.32\linewidth]{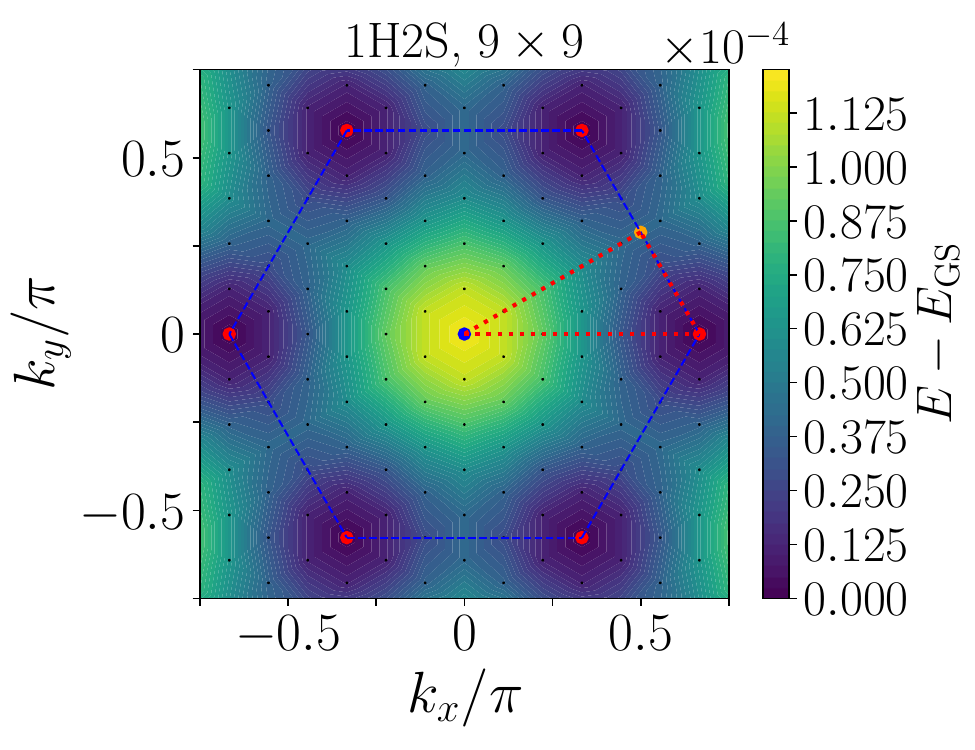}
    \includegraphics[width=0.32\linewidth]{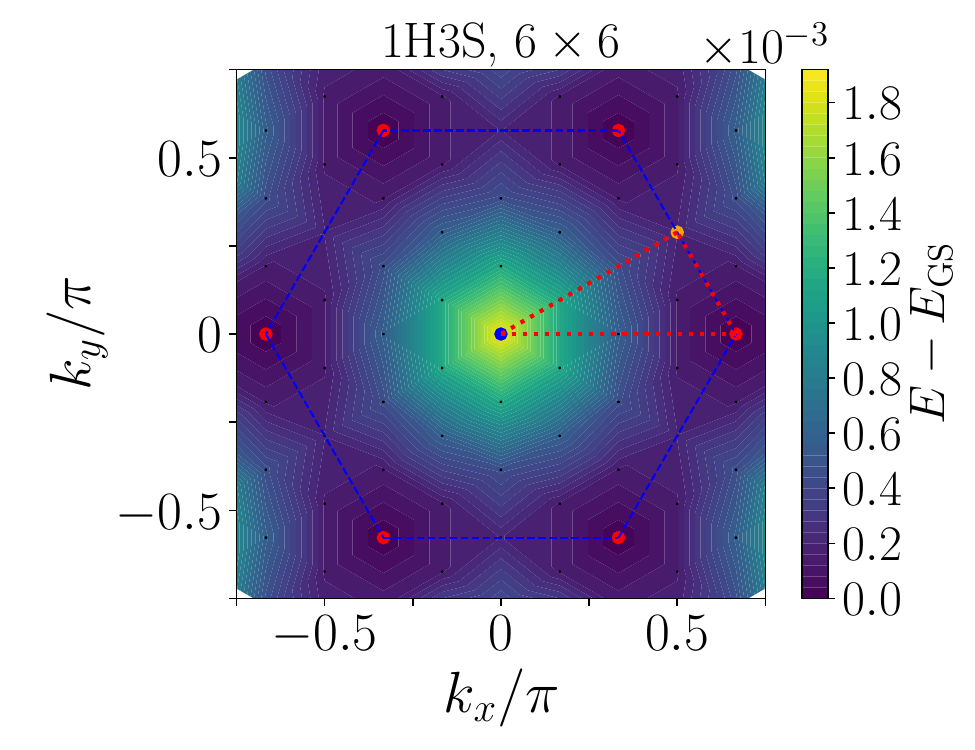}
    \includegraphics[width=0.32\linewidth]{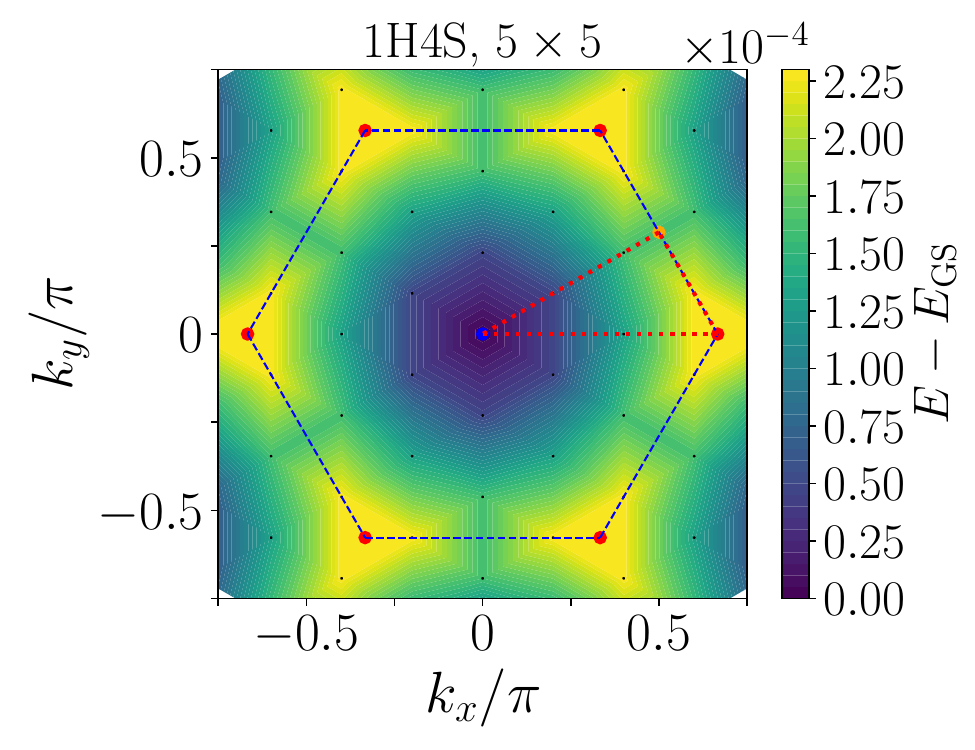}
    \hspace*{-1.2em}
    \includegraphics[width=0.32\linewidth]{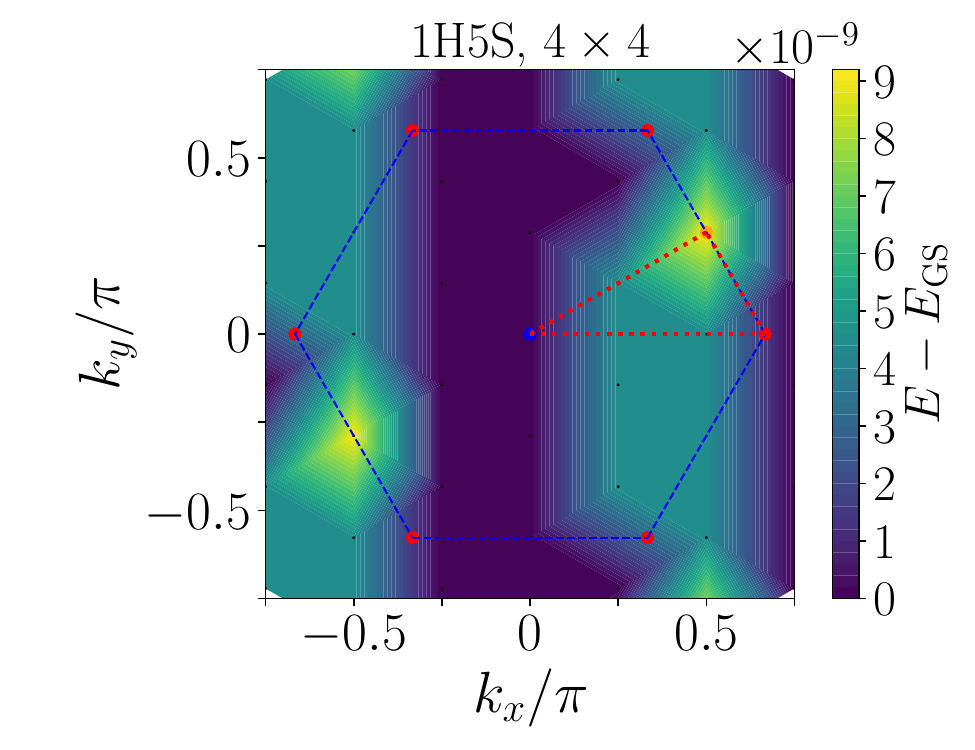}
    \hspace*{1.2em}
    \includegraphics[width=0.32\linewidth]{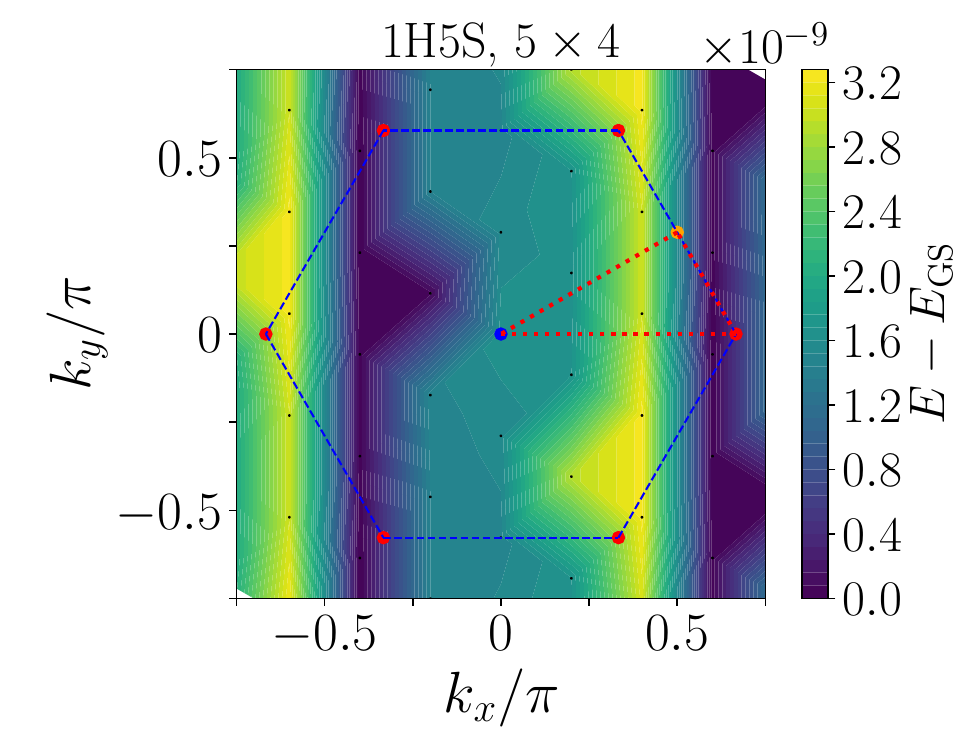}
    \caption{Brillouin zone plots for the lowest band in \hs{1}{n} systems, with $n=1, \dots, 5$. For $n=5$, we show results for both the $4\times4$ and $5\times4$ systems. We show in each of the plots the first BZ and the high symmetry points: $\Gamma$ (blue), $K$ (red) and $M$ (orange). The red dotted line indicates the path in reciprocal space used to plot Fig.~\ref{fig:Sband}.}
    \label{fig:SBZ}
\end{figure}

We present here additional results obtained using ED that further illustrate the behavior of our system. 
\begin{enumerate}
    \item The ground state hole-spin correlations and spin-spin correlations, computed after projecting the hole onto a given site, are shown in Fig.~\ref{fig:S1} for \hs{1}{n} systems with $2\le n \le 4$. The hole-spin correlation data were used to plot Fig.~\ref{fig:2c} in the main text. In the spin-spin correlations, we find that the two bonds nearest-neighboring the hole gradually approach the singlet limit ($-3/4$) as $n$ increases; we also see that the third-neighbor bonds become progressively more antiferromagnetic, which is consistent with the Husimi cactus picture (Fig.~\ref{fig:3} in main text).
    \item The lowest $6$ bands, for \hs{1}{n} systems with $2\le n \le 4$, are shown in Fig.~\ref{fig:Sband}. For $n=2$ we are able to fit the lowest three bands using a single-particle tight-binding model with nearest- and next-nearest-neighbor hopping $t_1=-1.783(4)\times10^{-3}$ and $t_2=6(4)\times10^{-6}$, including an energy offset $E_0 \simeq -2.95$ (see top-left panel of Fig.~\ref{fig:Sband}). For $n=3,4$ this approach does not yield good agreement, and a suitable effective tight-binding model likely requires farther-range hopping terms (not pursued in this work). 
    \item Additional numerical results for the bandwidths of the \hs{1}{4} system. For $n=4$, we performed ED on system sizes $3\times3$, $6\times3$, $4\times4$, $6\times4$, and $5\times5$, with the corresponding bandwidths listed in Table~\ref{tab:table}. The bandwidth obtained for the $3\times3$ system is substantially larger than those for the other system sizes and is therefore likely to be strongly affected by finite-size effects. A similar caveat applies to the $6\times3$ cluster, for which the band maximum appears at the $M$ point instead of the $K$ point, in contrast to the behavior suggested from larger system sizes. For the remaining three cluster sizes, the extracted bandwidths show some variation. Since the $6\times6$ system is beyond our current computational capabilities, we use the $5\times5$ result as our best estimate of the bandwidth and we use this value in the inset of Fig.~\ref{fig:2e} in the main text. We plot in Fig.~\ref{fig:Sbw1} all bandwidth data simulated in systems with $L_x>3$ and $L_y>3$. 
    \item Finite-size scaling analyses of the ground state energy, of the bandwidth of the bottom band, and of the hole-spin correlation for the \hs{1}{1} system are shown in Fig.~\ref{fig:Sf1} and~\ref{fig:Sf2}. We compare data obtained for $6\times6$, $9\times9$ and $12\times12$ systems and we observe rapid convergence, giving us confidence that finite size corrections are well under control in our results. 
    \item The energy of the lowest band, relative to the ground state energy, is shown across the full BZ for \hs{1}{n} systems with $n=1,\dots,5$ in Fig.~\ref{fig:SBZ}. Note the remarkably different scales of the color bars. For \hs{1}{5}, we present results obtained on both $4\times4$ and $5\times4$ systems, showing bandwidths of $9.10\times10^{-9}$ and $3.23\times10^{-9}$, respectively, with the latter used in the inset of Fig.~\ref{fig:2e} in the main text. These values are also plotted in Fig.~\ref{fig:Sbw1}. Together with the lack of structure reflecting the lattice symmetries, we are led to conclude that these features are likely within numerical error, and that our data are consistent with vanishing bandwidth and polaron localization. 
    \item The band structure for the unpolarized $3\times 3$ system is shown in the bottom-right panel of Fig.~\ref{fig:Sband}. The bandwidth of the bottom band is approximately $5\times10^{-3}$. Unfortunately the study of larger system sizes for unpolarized systems was not possible with our computational resources, and we are unable to assess how important finite-size effects may be in this case. 

\end{enumerate}

\begin{table}[bp]
\centering
\begin{tabular}{|c|c|c|c|c|}
\hline
$L_x$ & $L_y$ & Bandwidth                & Comments                                                                                                               \\ \hline
3     & 3     & $4.89\times10^{-4}$    & strong finite size effects expected                                                                                     \\ \hline
6     & 3     & $2.62\times10^{-4}$    & \begin{tabular}[c]{@{}c@{}}maximum of the band not at $K$, possibly due to \\ finite size effects along $y$-direction\end{tabular} \\ \hline
4     & 4     & $1.94\times10^{-4}$    &                                                                                                                        \\ \hline
6     & 4     & $2.29\times10^{-4}$    &                                                                                                                        \\ \hline
5     & 5     & $2.39\times10^{-4}$    & largest size accessible                                                                                                \\ \hline
\end{tabular}
\caption{Bandwidth data for the \hs{1}{4} system with different system sizes.}
\label{tab:table}
\end{table}
%
%

\section{Additional DMRG results}
We present here additional results obtained using DMRG that further illustrate the behavior of our system (and some were included in plots in the main text). 
%
%

\subsection{Localized wavefunctions in DMRG}
As already mentioned in the main text, our \hs{1}{0} system hosts a well-known flat band of compact-localized GSs~\cite{Mielke1992ka} (the fully polarized sector is equivalent to a single-particle tight-binding model). We also discussed how seemingly-localized GSs appear once again for $n\geq5$. In the latter case, an exact solution for these states is unknown to date, and it is therefore important to critically assess how reliably localization can be inferred from numerics alone -- which we shall discuss in detail below. We note however that, leaving aside the conceptual question of localized vs. extremely massive polarons, one point that emerges robustly from our study is that the mass of the polarons increases abruptly by at least $6$ orders of magnitude between $n<5$ and $n\geq5$. 

Let us use the \hs{1}{0} system for reference, with its lowest band being flat at $E=-2$ and consisting of states compact-localized on individual hexagonal plaquettes. 
A priori, numerical algorithms are free to select arbitrary superpositions of such states. One can then add a small energetic bias (e.g., a chemical potential around a given plaquette) to select a given compact localized GS. In particular, when using DMRG we find that a small bias is not sufficient, likely due to the flatness of the band, and one has to apply a chemical potential $\gtrsim0.1$. 

\begin{figure}
    \centering
    \includegraphics[width=0.32\linewidth]{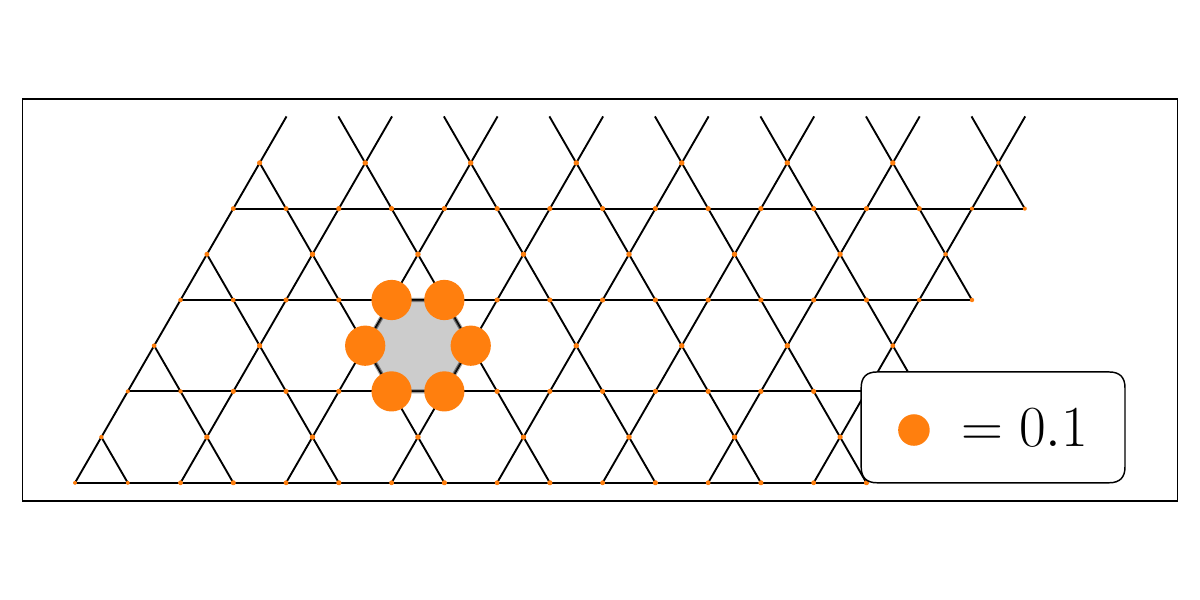}
    \includegraphics[width=0.32\linewidth]{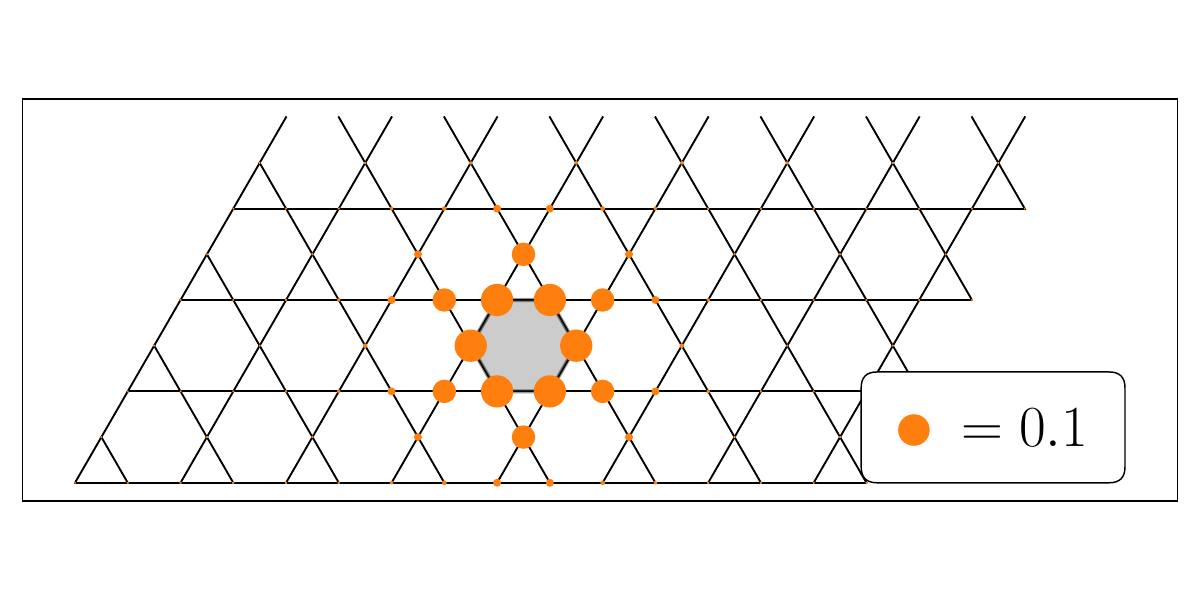}
    \includegraphics[width=0.32\linewidth]{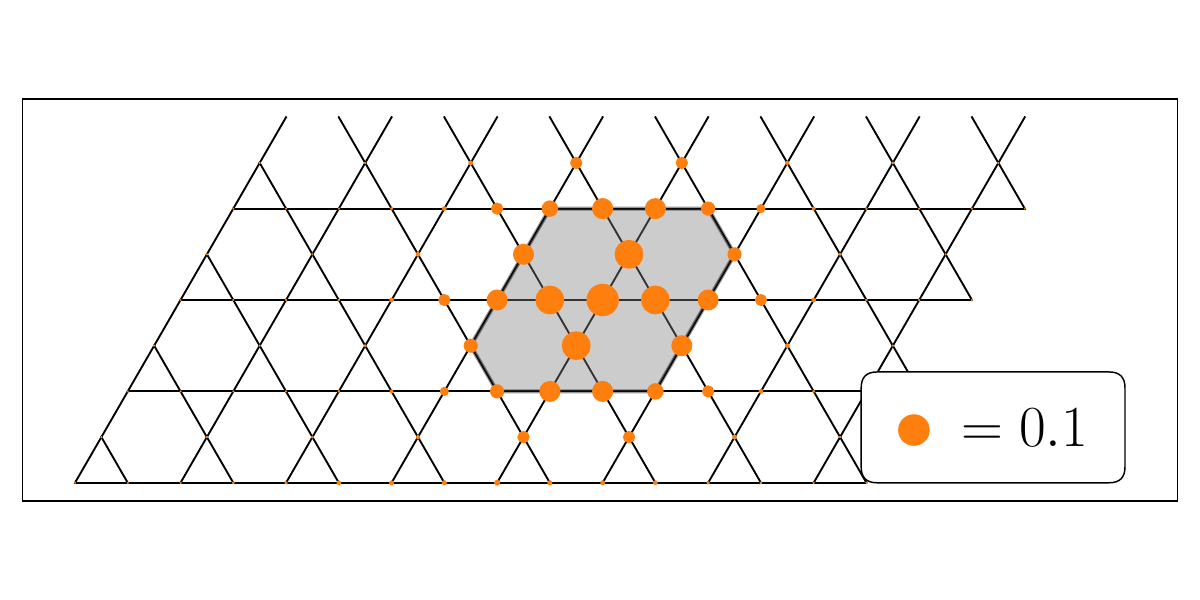}
    \caption{DMRG results for \hs{1}{0}, \hs{1}{5} and \hs{1}{12}. Pinning potentials are initially added to the sites on the shaded area and subsequently removed. For \hs{1}{5} and \hs{1}{12}, these results are in quantitative agreement with those obtained from unbiased DMRG simulations (Fig.~\ref{fig:84DMRG}).}
    \label{fig:SpinDMRG}
\end{figure}

To our surprise, for \hs{1}{n} systems with $n\geq5$, we find instead that DMRG is able to obtain what look like localized states even in the absence of a bias breaking their translational symmetry on the lattice. For comparison, we run DMRG simulations where we applied a chemical potential of order $0.1$ to the sites corresponding to the main support of one of these localized states, and then adiabatically reduced the bias to zero. We verified that the final state is in good quantitative agreement with the state obtained from DMRG run without the bias altogether (e.g., by comparing hole and spin densities) -- see Fig~\ref{fig:SpinDMRG} for results with \hs{1}{0}, \hs{1}{5} and \hs{1}{12}, where suitably selected pinning sites successfully find the correct localized state. 

%
%

\subsection{Polarized sectors}
\begin{figure}
    \centering
    \includegraphics[width=0.32\linewidth]{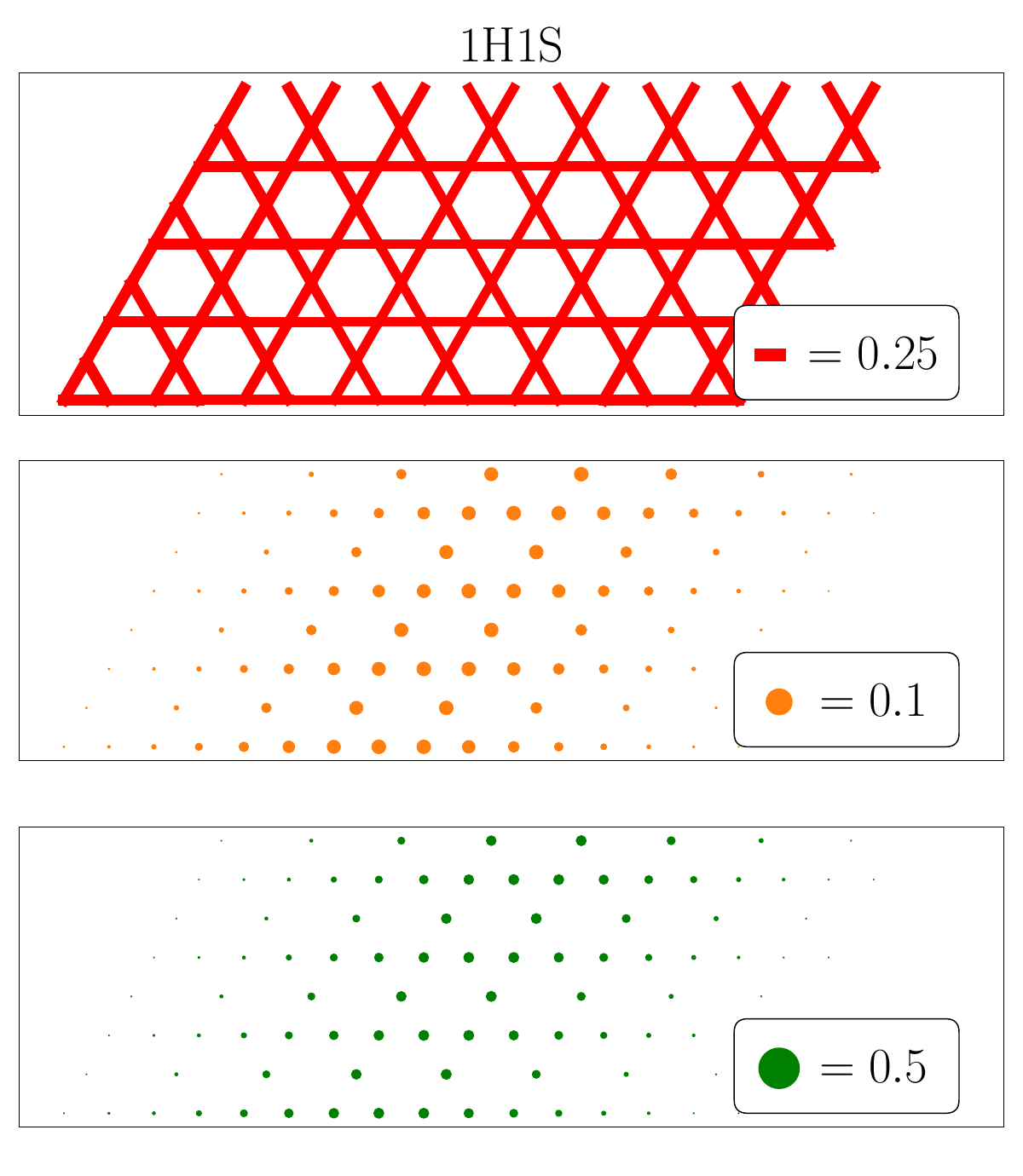}
    \includegraphics[width=0.32\linewidth]{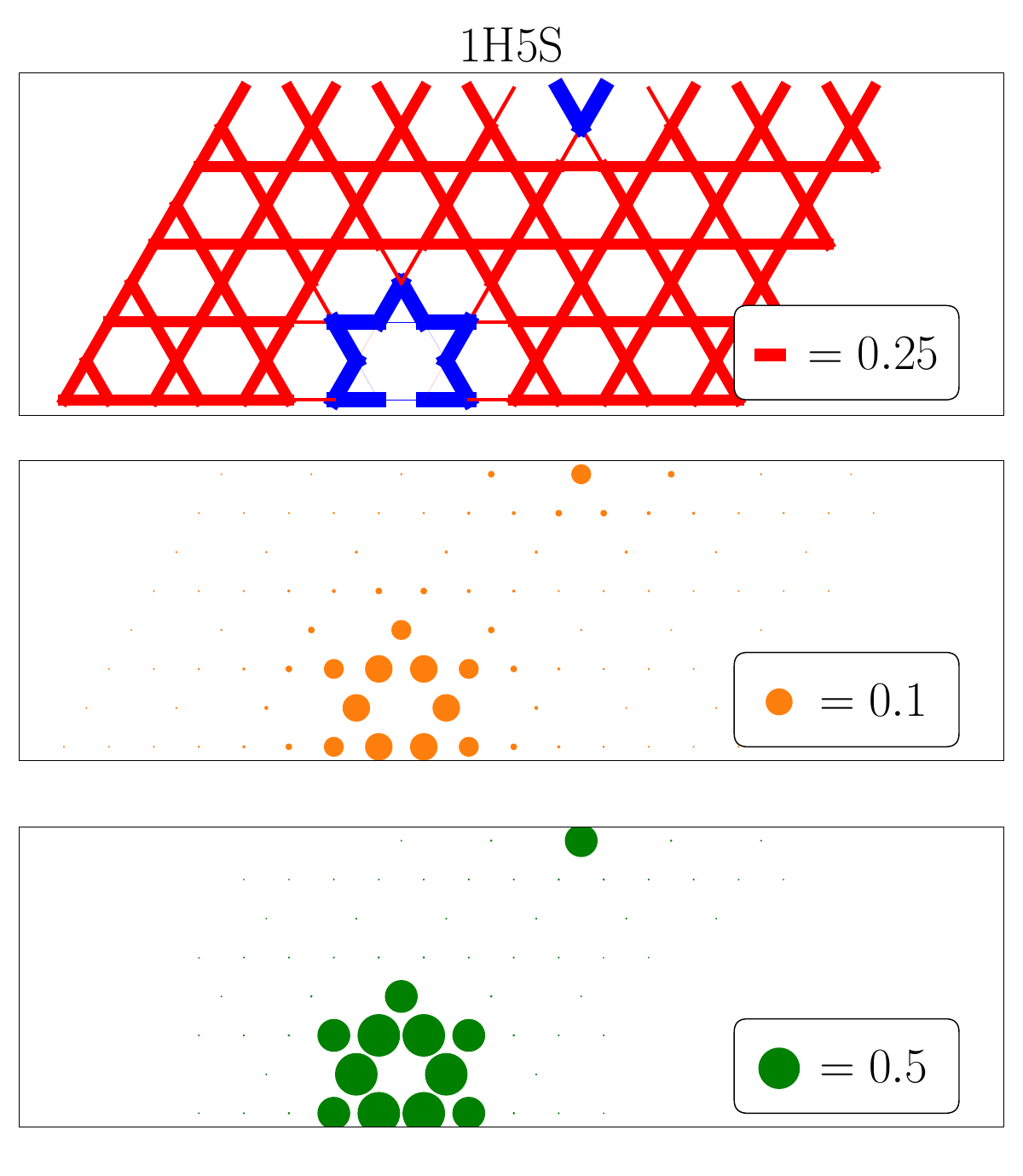}
    \includegraphics[width=0.32\linewidth]{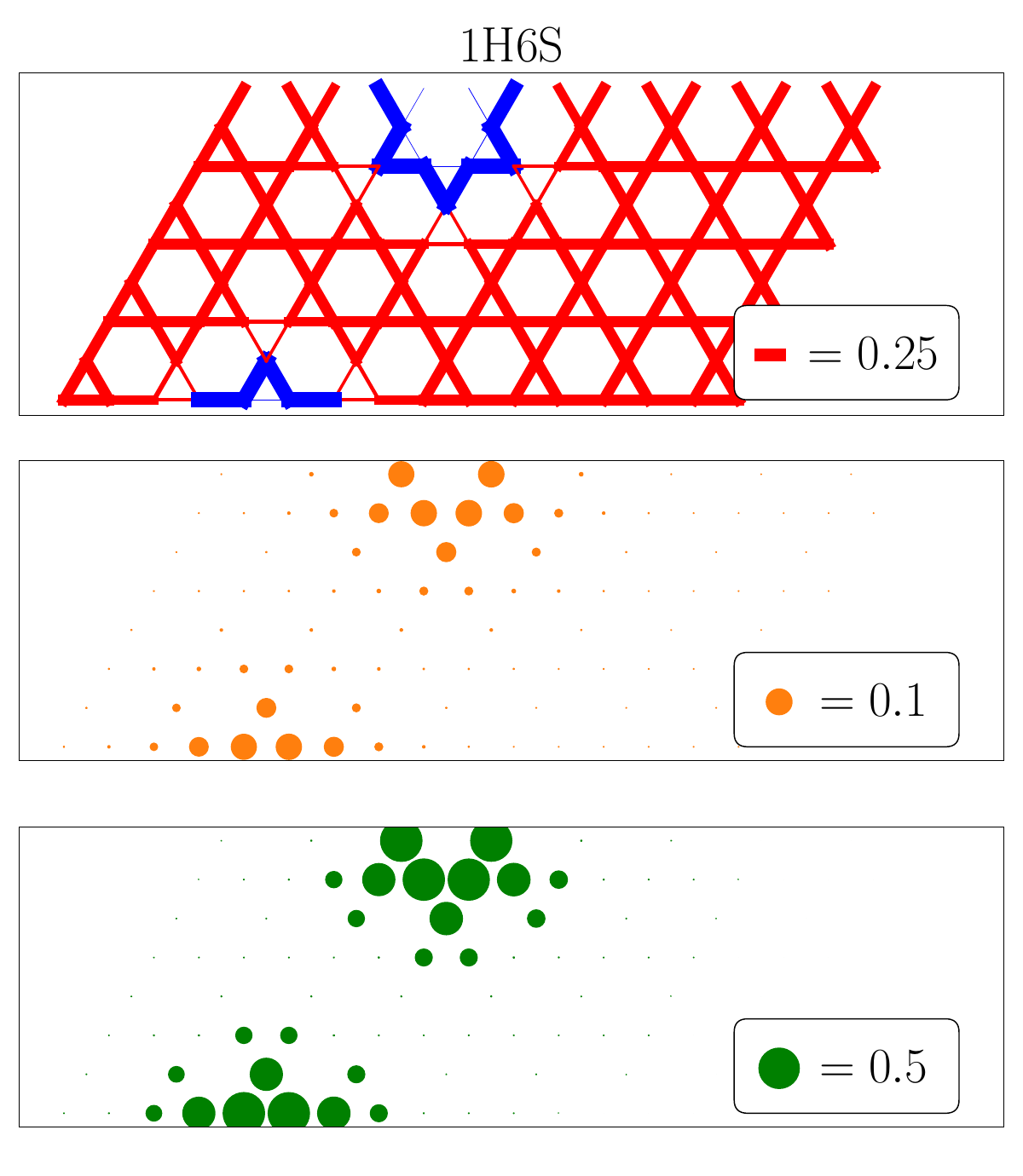}
    \includegraphics[width=0.32\linewidth]{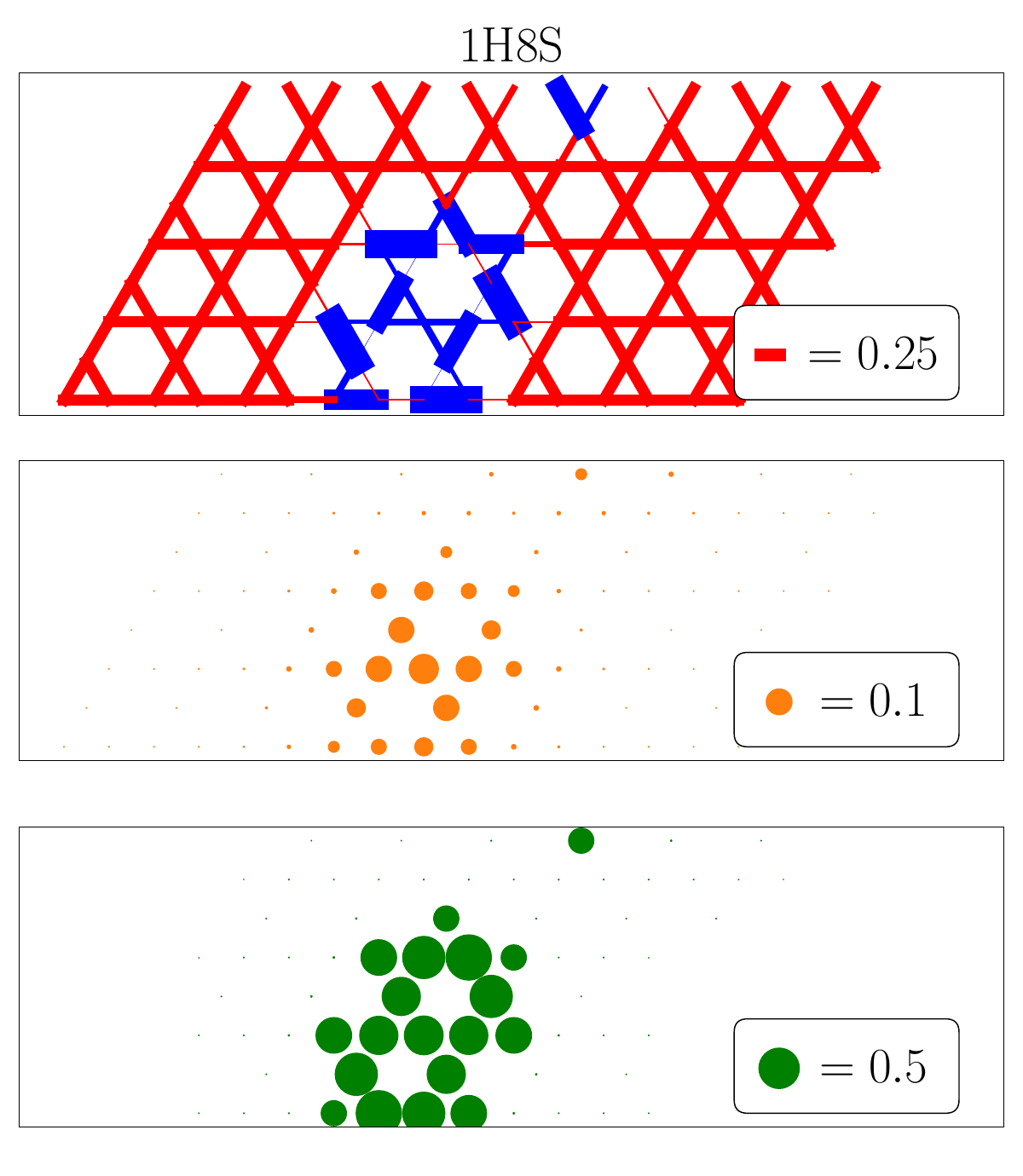}
    \includegraphics[width=0.32\linewidth]{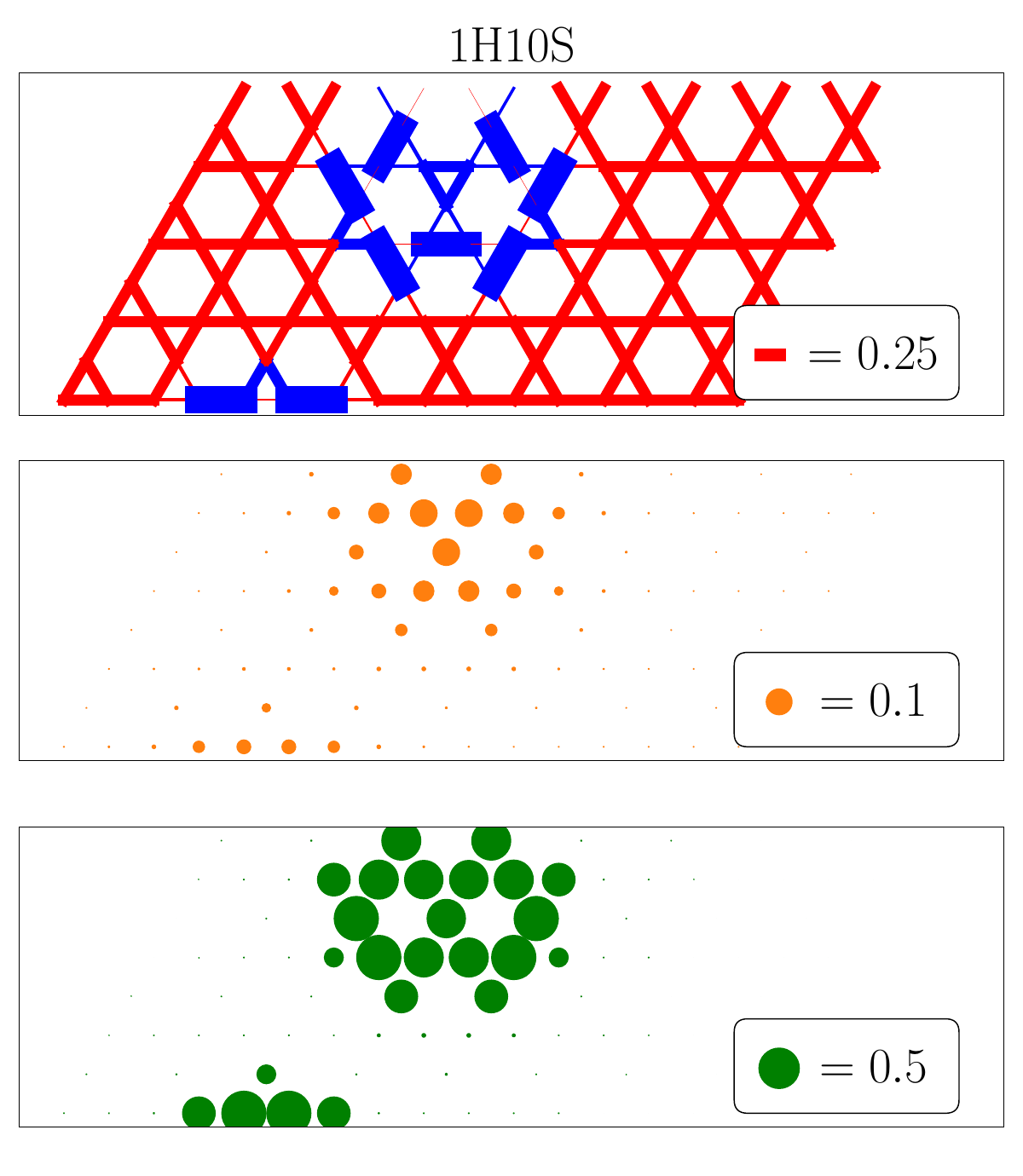}
    \includegraphics[width=0.32\linewidth]{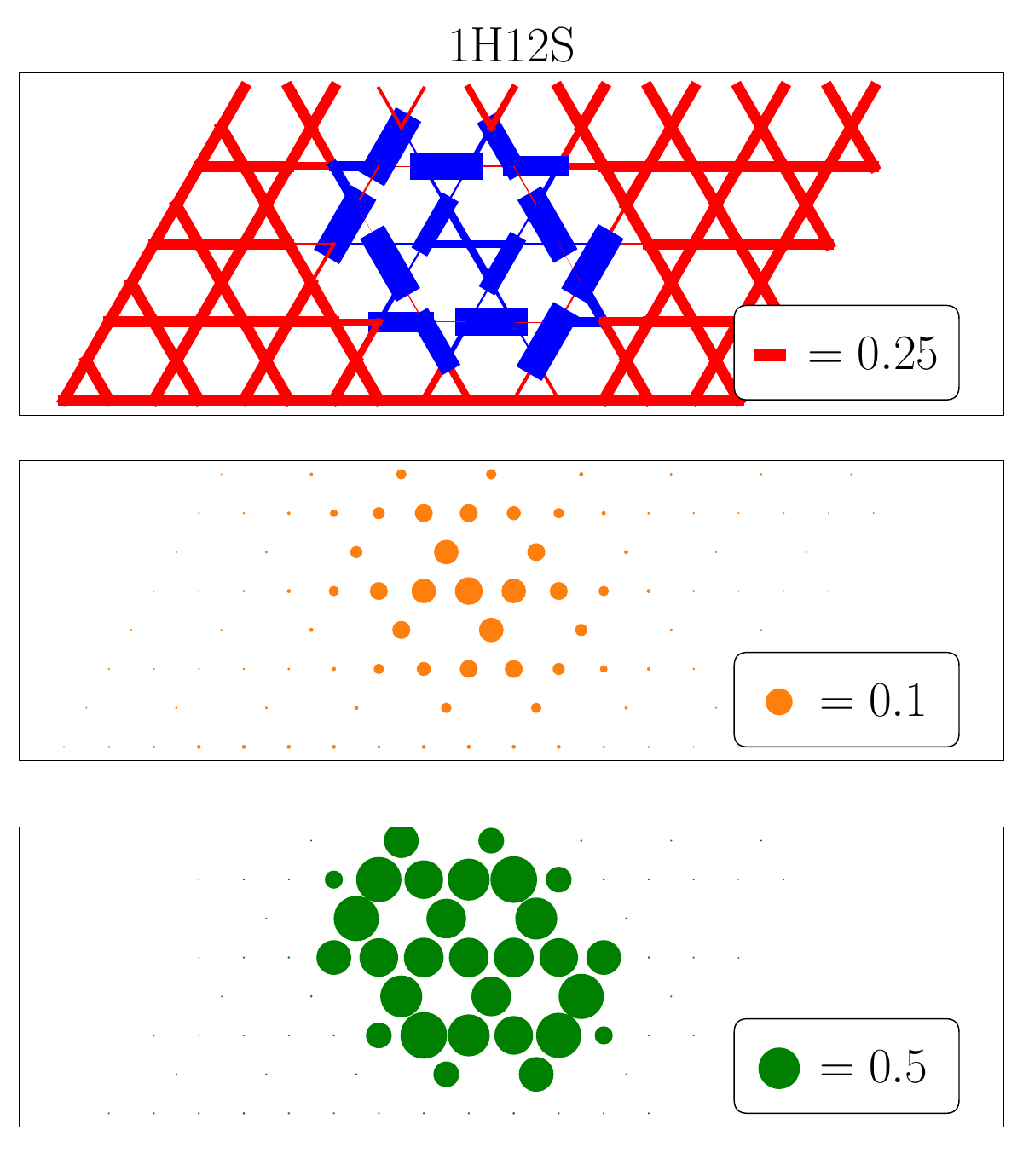}
    
    \caption{DMRG data obtained for $8\times4$ systems, with periodic boundary conditions along the $y$-axis and open boundary conditions along the $x$-axis. We show results obtained for the \hs{1}{n} systems, with $n = 1, 5, 6, 8, 10, 12$: spin-spin correlation function (top), hole density (middle) and reversed-spin density (bottom). For all configurations with $n\ge5$, we observe seemingly-localized polarons with increasingly larger size. For some configurations, translation along the $y$-axis was performed to visually center the polaron in the figure.}
    \label{fig:84DMRG}
\end{figure}

\begin{figure}
    \centering
    \includegraphics[width=0.32\linewidth]{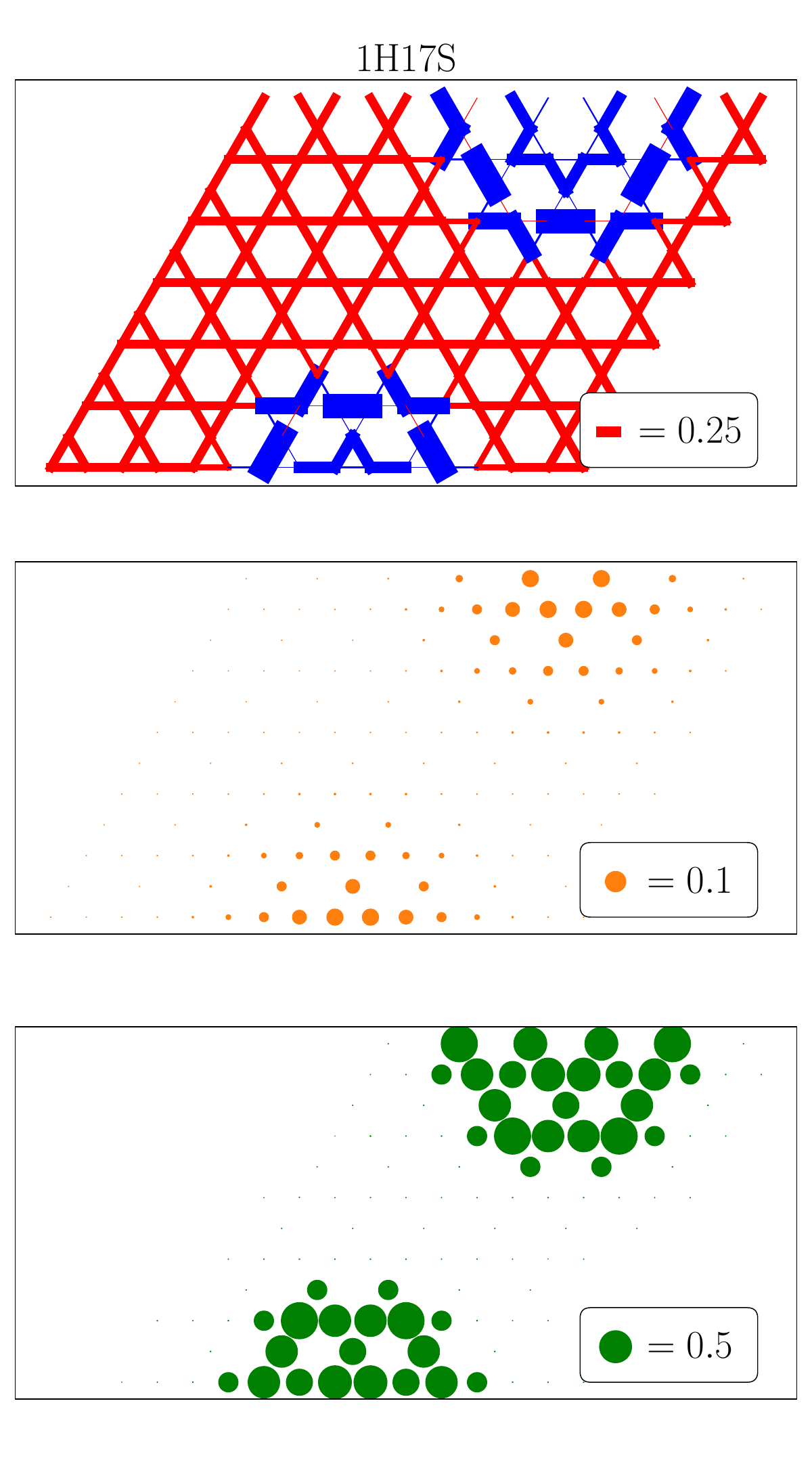}
    \includegraphics[width=0.32\linewidth]{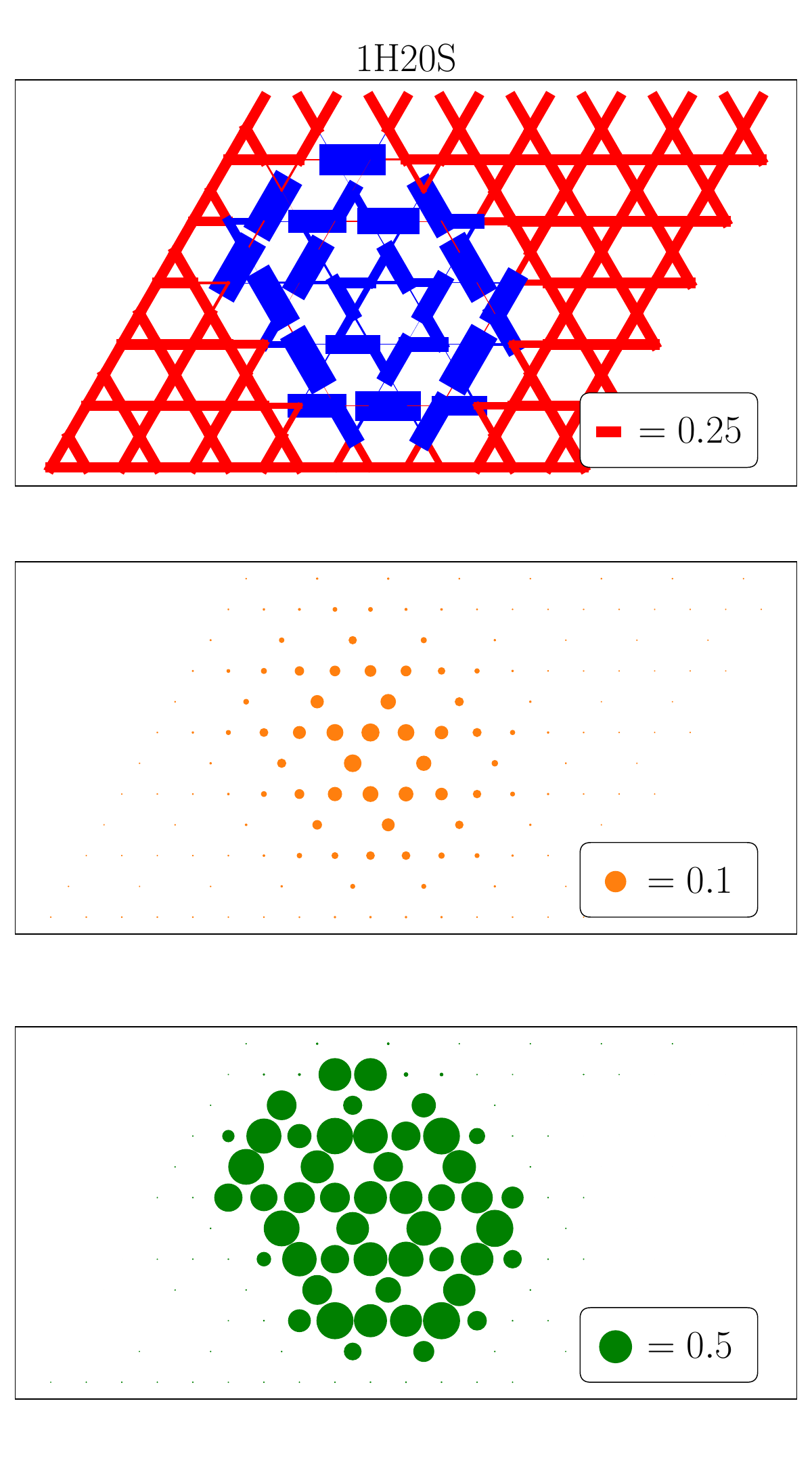}
    \includegraphics[width=0.32\linewidth]{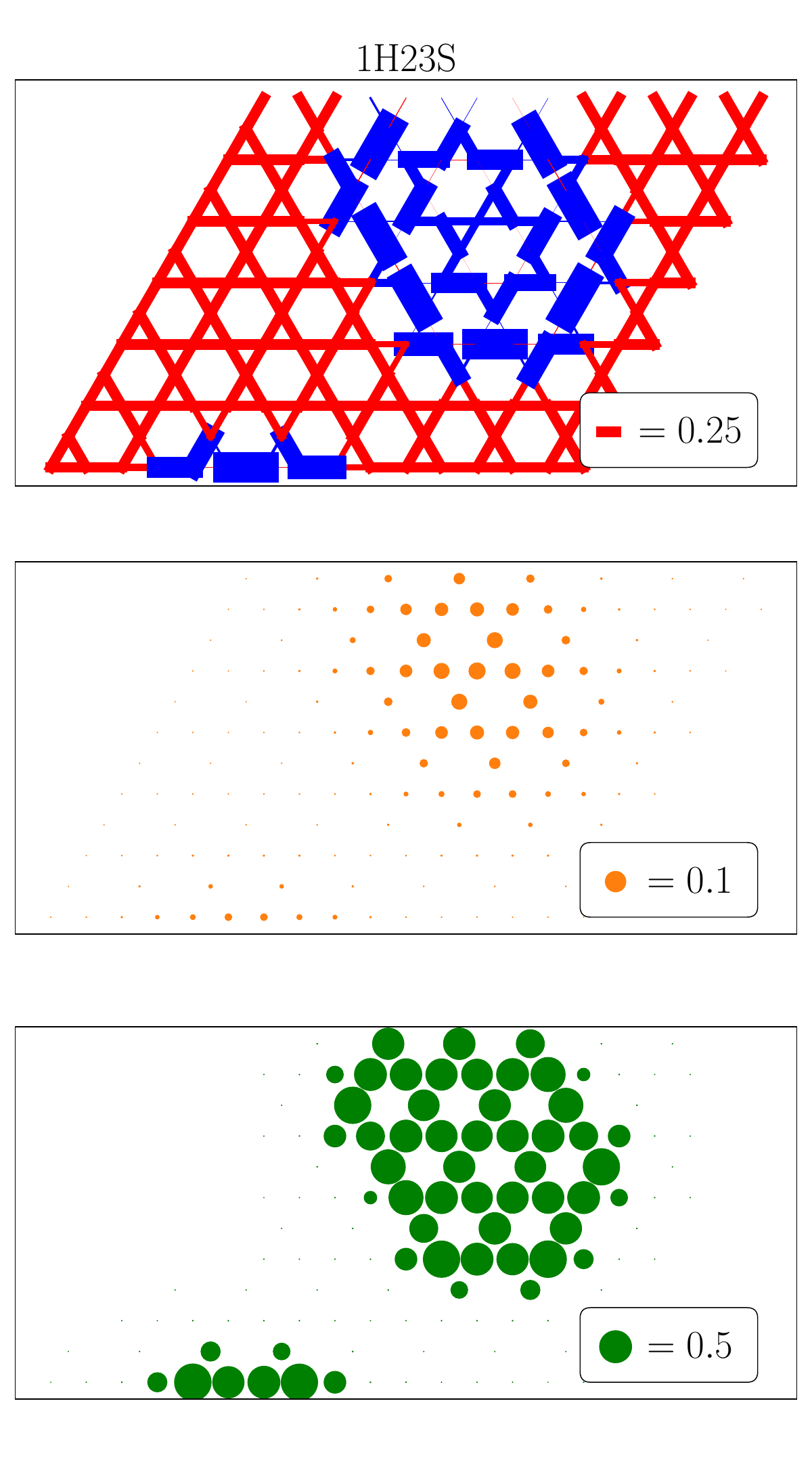}
    
    \caption{DMRG data obtained for the $8\times6$ systems, with periodic boundary conditions along the $y$-axis and open boundary conditions along the $x$-axis. We show results obtained for $n=17, 20, 23$. The panels are arranged as in Fig.~\ref{fig:84DMRG}.}
    \label{fig:86DMRG}
\end{figure}

We present here additional DMRG results as a function of magnetization sector. In Fig.~\ref{fig:84DMRG} we show DMRG results for \hs{1}{n}, on $8\times4$ YC-geometries with $n=1, 5, 6, 8, 10, 12$: nearest-neighbor spin-spin correlation (top), hole density (middle) and reversed-spin density (bottom). We also show DMRG results on $8\times6$ geometries with $n=17, 20, 23$ in Fig.~\ref{fig:86DMRG}. The energies of these states are plotted in Fig.~\ref{fig:2d} in the main text.

\begin{figure}
    \centering
    \includegraphics[width=0.24\linewidth]{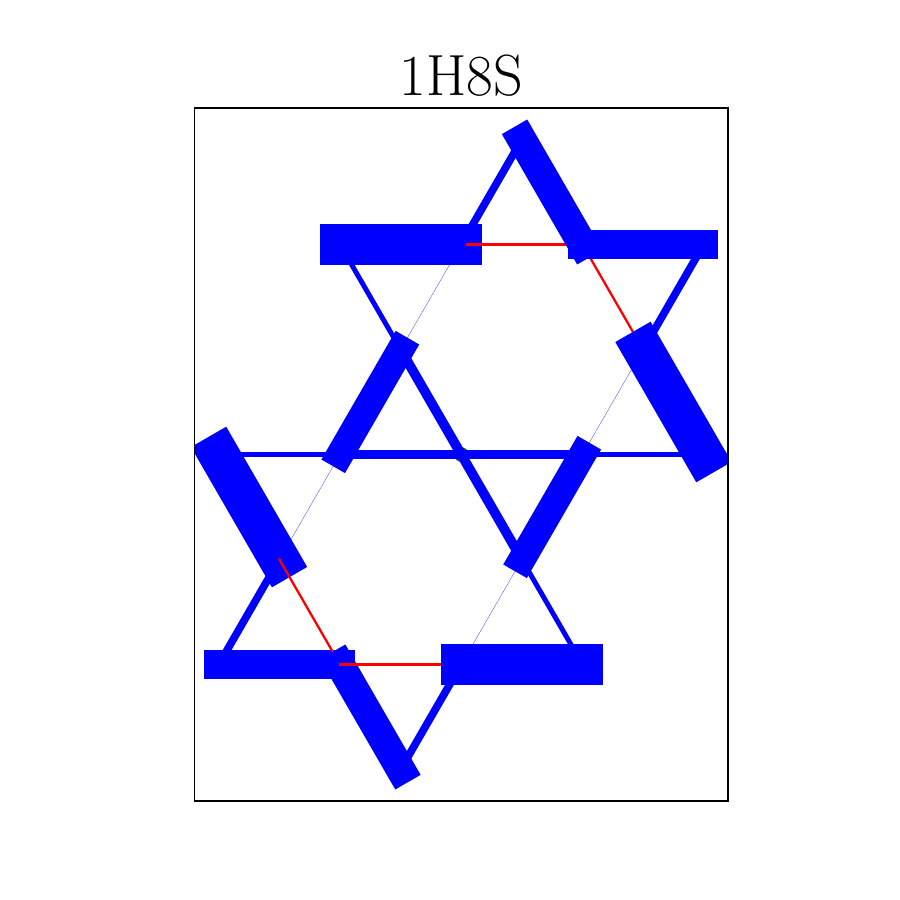}
    \includegraphics[width=0.24\linewidth]{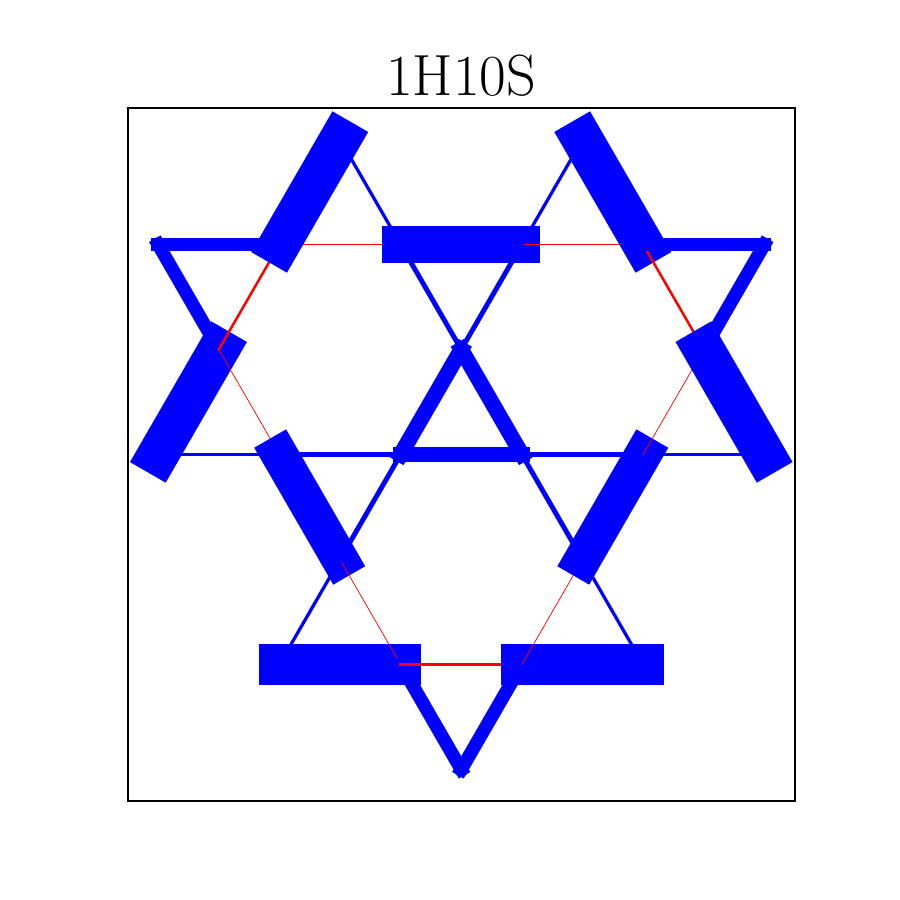}
    \caption{ED results on clusters corresponding to the \hs{1}{n} polaron, with $n=8, 10$ respectively. The spin correlation functions are quantitatively similar to those obtained using DMRG in Fig.~\ref{fig:84DMRG}.} 
    \label{fig:SclusterED}
\end{figure}

For $n=8$ and $10$, similarly to the $n=5$ case discussed in the main text, we are able to make sense of the polaron shape by performing ED on corresponding clusters consisting of $19$ and $24$ sites, respectively, as shown in Fig.~\ref{fig:SclusterED}. We are also able to provide variational wavefunctions using our Husimi cactus approach (see main text): we identify all compatible Husimi clusters obtained by removing triangles from the ED cluster and checking that the resulting shape is a connected Husimi cactus; we compute the exact ground state $|\psi_\mathrm{HC}^{(i)}\rangle$ for each of them; we then search for a variational state with the lowest energy in the subspace spanned by these states, $|\psi_\mathrm{HC}\rangle=\sum_{i}\alpha_i|\psi_\mathrm{HC}^{(i)}\rangle$, by solving the generalized eigenvalue problem $\hat{A}\bm{\alpha}=E\hat{B}\bm{\alpha}$ with $\hat{A},\hat{B}$ being matrices with $\hat{A}_{ij}=\langle\psi_\mathrm{HC}^{(i)}|\mathcal{H}|\psi_\mathrm{HC}^{(j)}\rangle$ and $\hat{B}_{ij}=\langle\psi_\mathrm{HC}^{(i)}|\psi_\mathrm{HC}^{(j)}\rangle$. We find a $98\%$ and $97\%$ overlap, 
respectively. 
Similar calculations could in principle be done for the \hs{1}{12} system, involving ED performed on a $29$-site cluster, but these are beyond our computing capability.

The variational approach becomes progressively more convoluted as the size of the polaron grows (involving $6$, $20$, $63$ and $242$ configurations for $n=5, 8, 10, 12$), making the effective modeling less straightforward to interpret. 
We note that the \hs{1}{17} spin-spin correlator is reminiscent of the $36$-site unit cell order proposed in Ref.~\onlinecite{singh2007ground} for the quantum spin-$1/2$ Heisenberg model on the {\kag} lattice, suggesting an interesting direction for future investigation. 

%
%

\subsection{Unpolarized sector}

\begin{figure}
    \centering
    \subfloat[]{
        \centering
        \includegraphics[width=0.295\linewidth]{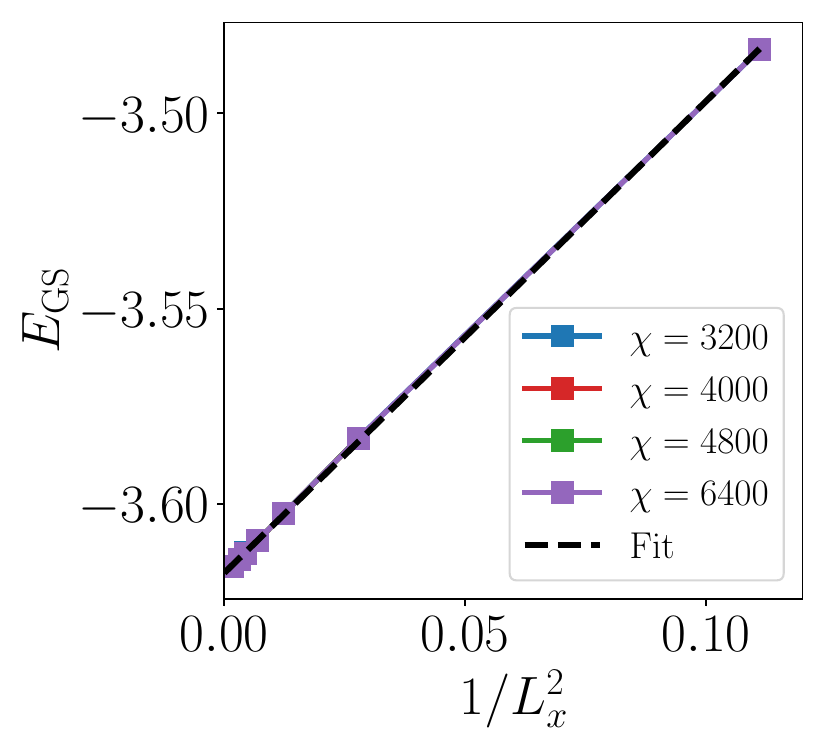}
        \label{fig:Sunp_E}
    }
    \subfloat[]{
        \centering
        \raisebox{0.em}{
        \includegraphics[width=0.36\linewidth]{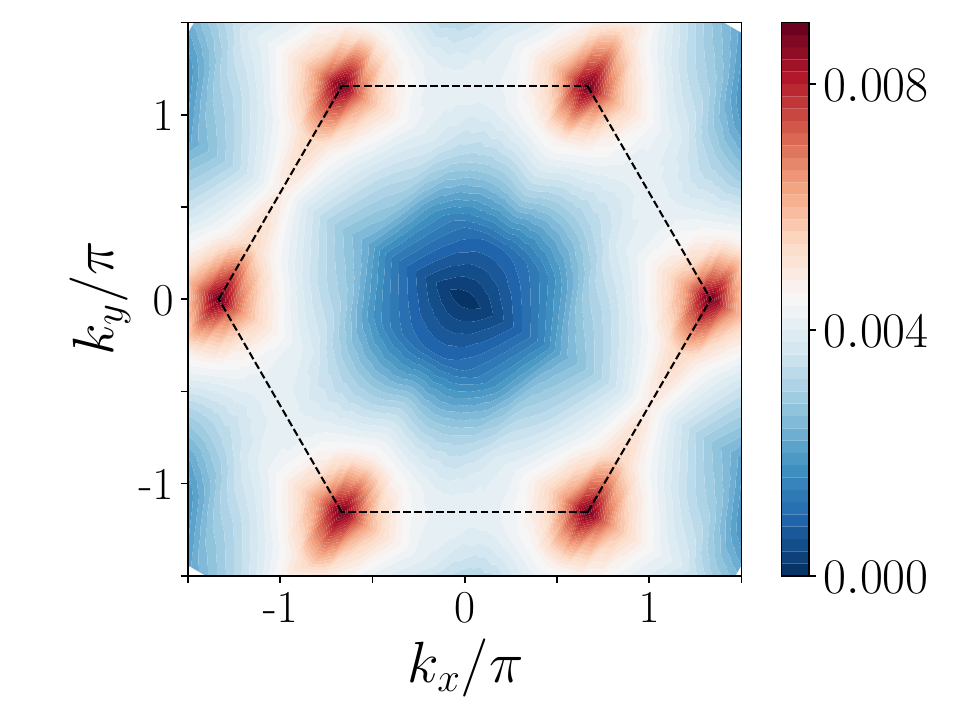}
        }
        \label{fig:Sunp_Sk}
    }
    \subfloat[]{
        \centering
        \raisebox{0.53em}{
        \includegraphics[width=0.31\linewidth]{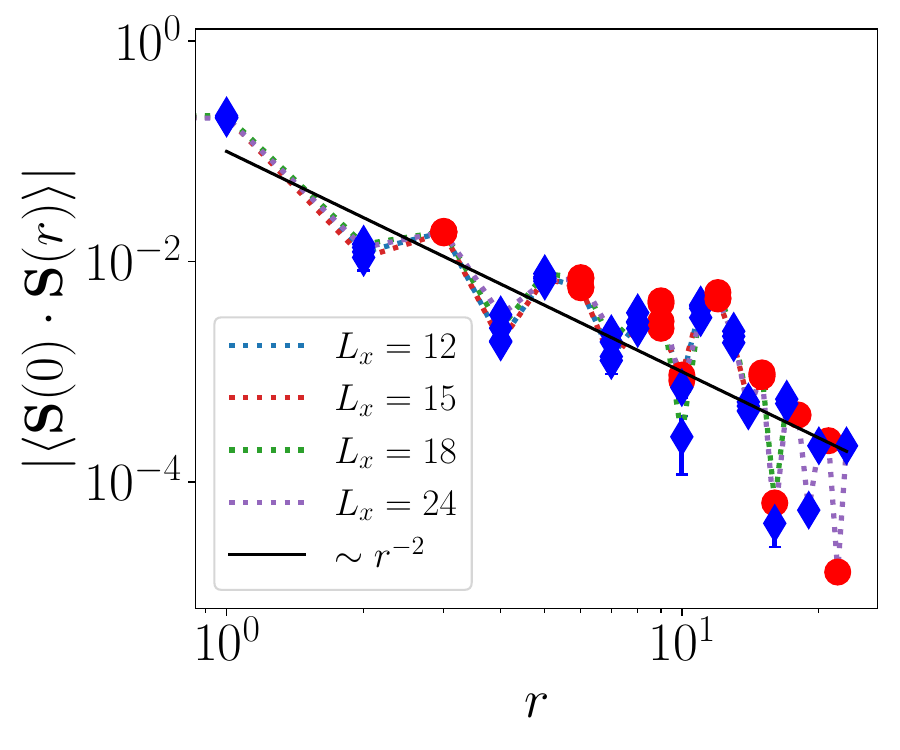}
        }
        \label{fig:Sunp_Sx}
    }
    \caption{(a) GS energies $E_\mathrm{GS}$ obtained using DMRG simulations for system sizes $L_x \times 3$, with $3\le L_x \le 24$. Linear scaling $E_\mathrm{GS} \propto 1/L_x^2$ is observed, which results in an extrapolated value $E_\mathrm{GS}=-3.618(5)$. 
    (b) Spin structure factor $S(\mathbf{k})$ in the extended BZ scheme, with clear peaks at the $K$ points, for system size $24\times3$ with $\chi=6400$. 
    (c) Absolute values of the spin correlations along the $x$-axis for different system sizes. Red (blue) points indicate positive (negative) correlators. Extrapolation for $\chi\to\infty$ was performed for each system size. A $\sim r^{-2}$ decay is plotted (black solid line) as a guide to the eye. 
    }
    \label{fig:Sunp_corr}
\end{figure}

\begin{figure}
    \centering
    \includegraphics[width=\linewidth]{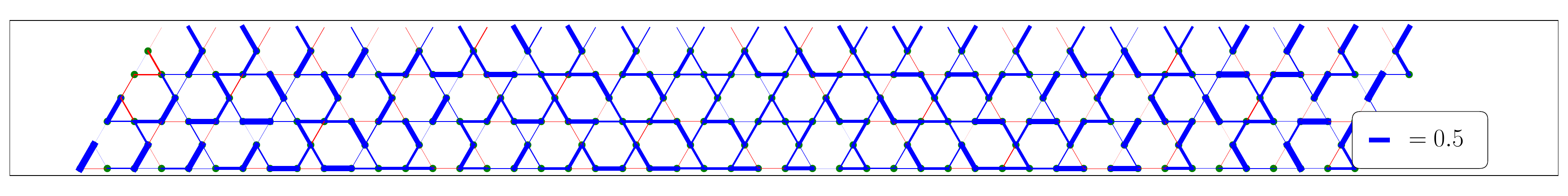}
    \includegraphics[width=\linewidth]{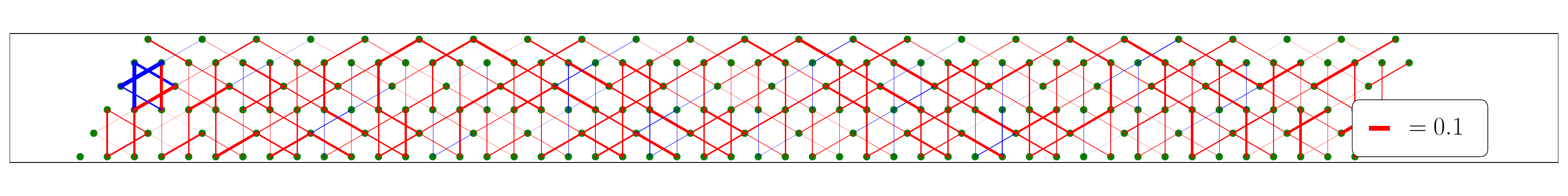}
    \includegraphics[width=\linewidth]{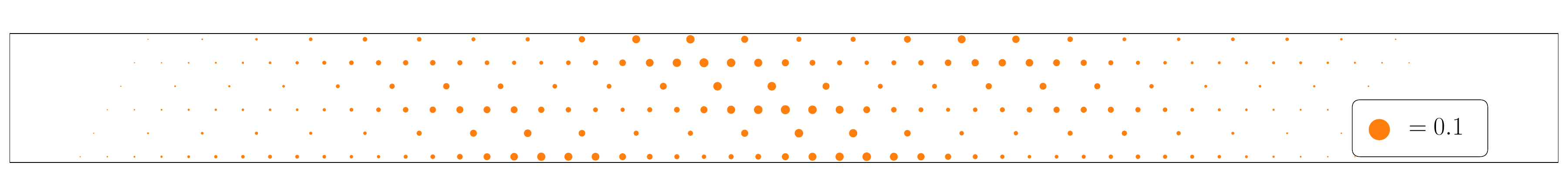}
    \caption{DMRG data obtained for the unpolarized sector on the $24\times3$ lattice, with periodic boundaries in the $y$-direction and open boundaries in the $x$-direction. Both nearest-neighbor and next-nearest-neighbor spin correlations are shown (top and middle panels, respectively), together with the hole density (bottom panel).}
    \label{fig:Sunp}
\end{figure}

We present here the DMRG results obtained for the unpolarized system. The GS energy for various system sizes is shown in Fig.~\ref{fig:Sunp_E}, exhibiting linear scaling with $1/L_x^2$. From this scaling, we extrapolate a value $E_{\rm{GS}}=-3.618(5)$ for quasi-one-dimensional systems with $L_y=3$. This value is unlikely to be close to the true GS energy of the infinite two-dimensional system, due to significant self-interaction effects in the $y$-direction. The result is nonetheless consistent with the extrapolated energies obtained in the polarized sectors (see Fig.~\ref{fig:2d} in the main text), where a value of $E\approx -3.68$ is found. 

In Fig.~\ref{fig:Sunp} we report results obtained for a system of dimensions $L_y=3,\,L_x=24$ and bond dimension $\chi=6400$, for several local observables. The spin-spin correlations for nearest-neighbor pairs appear to be predominantly antiferromagnetic, whereas for next-nearest-neighbor pairs they appear to be predominantly ferromagnetic. This pattern is consistent with the presence of {\rord} order in the system (see Fig.~\ref{fig:Spotts3}). 

We further compute the static spin structure factor $S(\mathbf{k})$ over the entire (expanded) first Brillouin zone (BZ), illustrated in Fig.~\ref{fig:Sunp_Sk}. The pronounced peaks at the $K$ points are characteristic of {\rord} order, whereas the alternative $q=0$ order (see Fig.~\ref{fig:Spotts0}) would generate dominant peaks at the $M$ points (or, equivalently, at the $\Gamma$ points of the second, unexpanded BZs) -- which are not observed. 

As an additional indicator of {\rord} order, we present in Fig.~\ref{fig:Sunp_Sx} the absolute value of the spin-spin correlation functions along the $x$-direction obtained via DMRG. Each data point in this figure results from an average throughout the system and extrapolation in the bond dimension, performed by fitting the correlation values computed at different $\chi$ as a function of $1/\chi$ and extracting the $\chi\to\infty$ limit. 
{The sign of the correlators (represented by the choice of color for each data point) oscillates at short distance with period $3$, also consistent with {\rord} order. 
We shall compare these results with the correlators of the classical Potts models in the next section (Fig.~\ref{fig:SPottscorr}).} 

%
%

\section{Additional classical Monte-Carlo results for the Potts model on the {\kag} lattice} 

\begin{figure}
    \centering
    \subfloat[]{
        \centering
        \includegraphics[width=0.32\linewidth]{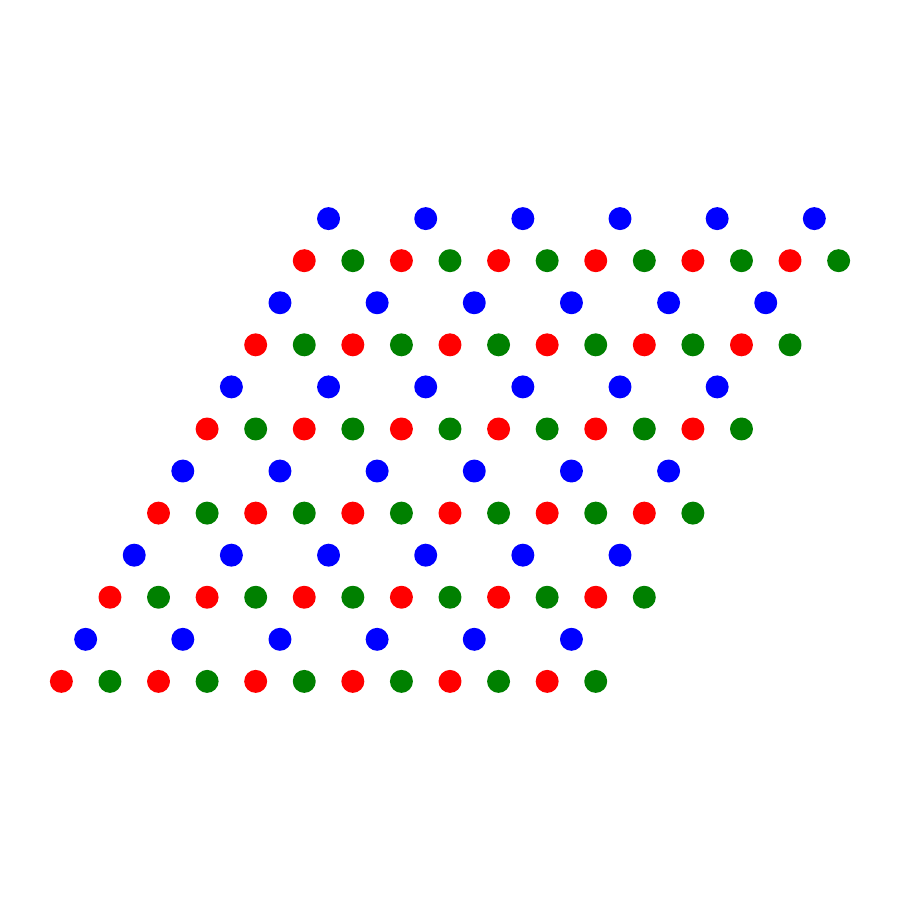}
        \label{fig:Spotts0}
    }
    \subfloat[]{
        \centering
        \includegraphics[width=0.32\linewidth]{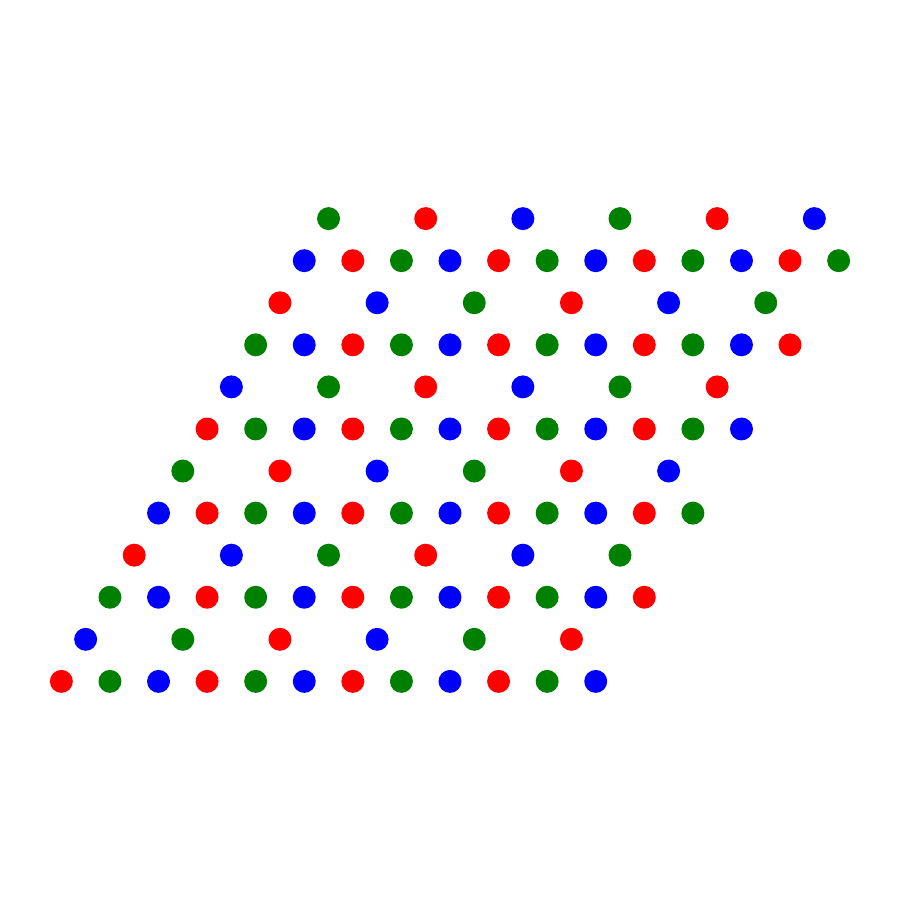}
        \label{fig:Spotts3}
    }
    
    \caption{Illustration of (a) the $q=0$ and (b) the {\rord} order for the $3$-state Potts model on the {\kag} lattice.}
    \label{fig:Spotts}
\end{figure}

\begin{figure}
    \centering
    \subfloat[]{
        \centering
        \includegraphics[width=0.39\linewidth]{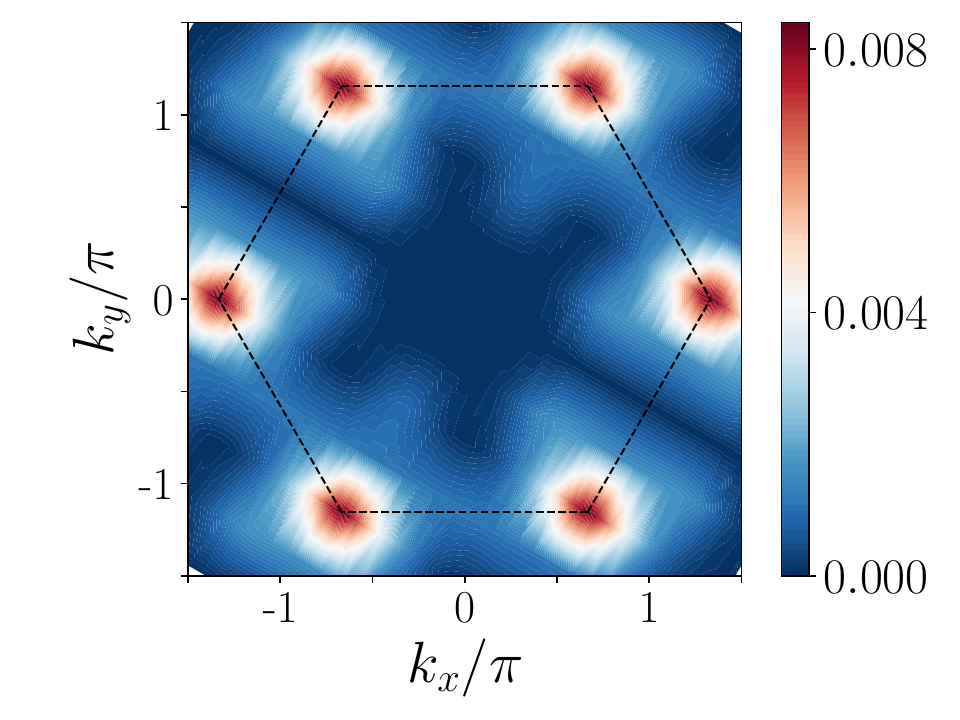}
        \label{fig:SPotts_k_corr}
    }
    \subfloat[]{
        \centering
        \raisebox{0.1em}{
            \includegraphics[width=0.35\linewidth]{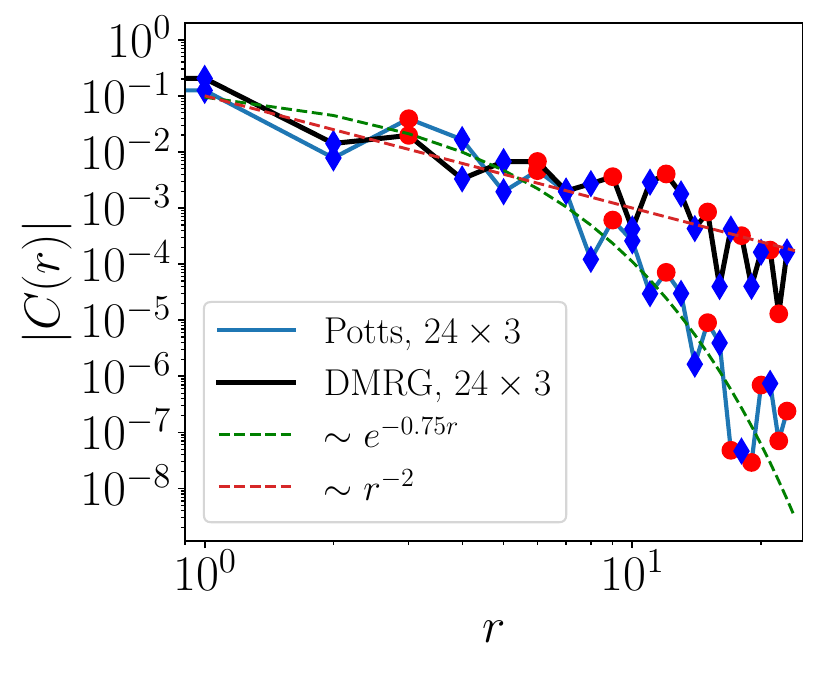}
        }
        \label{fig:SPotts_SzSz_corr}
    }
    \caption{(a) Intensity plot for $S(\mathbf{k})$ for the Potts model, showing  peaks at the $K$ points, for system size $24\times3$. 
    (b) Comparisons of spin correlation functions between the Potts model and our model on the {\kag} lattice. The same convention as in Fig.~\ref{fig:Sunp_Sx} is used. Exponential and power-law dashed lines are shown as guides to the eye.} 
    \label{fig:SPottscorr}
\end{figure}

The $3$-state Potts model on the {\kag} lattice hosts extensively degenerate ground states at zero temperature, with their number scaling approximately as $1.13^N$~\cite{baxter1970colorings}. 
Two examples are the $q=0$ state, characterized by preserving the original $3$-site unit cell (Fig.~\ref{fig:Spotts0}), and the {\rord} state, with an enlarged $9$-site unit cell (Fig.~\ref{fig:Spotts3}). 
Plentiful allowed local state re-arrangements afforded by the {\rord} state favor correlations with its characteristic wavevector when the average over all permitted states is taken. However, this does not lead to long-range order; instead, in the thermodynamic limit, one finds algebraically decaying correlations and the scaling of $S(\mathbf{k}=K)$ with system size reflects the exponent of these correlations at a Kosterlitz-Thouless phase transition. The utility of the comparison with the Potts model is that a strengthening of the correlations (in the sense of an increased stiffness in the appropriate height model description \cite{huse1992classical,chern2013dipolar}) implies the presence of long-range spin order; while a weakening corresponds to moving inside the KT phase, resulting in a drifting exponent of the algebraic correlations. 

We plot $S(\mathbf{k})$ in the entire first extended BZ of the Potts model simulated on a $24\times3$ system with cylindrical geometry in Fig.~\ref{fig:SPotts_k_corr}. Peaks at the $K$ points are clearly seen. Comparing with Fig.~\ref{fig:Sunp_Sk}, the peaks appear to have similar intensity. 

We further plot the correlation functions of both models in Fig.~\ref{fig:SPotts_SzSz_corr}. 
In our quasi-1D cylindrical geometry, we find that the correlations of the Potts model are weaker than in the model we consider. This suggests the possibility of long-range semiclassical \rord\ spin correlations in our model. However, 
due to limitations in DMRG computational resources, we are not able to extend such comparison to a conclusive finite-size study of two-dimensional systems. 
%
%

\section{12-site sawtooth ring system}
It is interesting to notice that the $12$-site system discussed in the main text for the \hs{1}{5} case can be thought of as a sawtooth chain of length $6$ with periodic boundary conditions~\cite{Glittum2026}. 

\begin{figure}[t]
    \centering
    \includegraphics[width=0.5\linewidth]{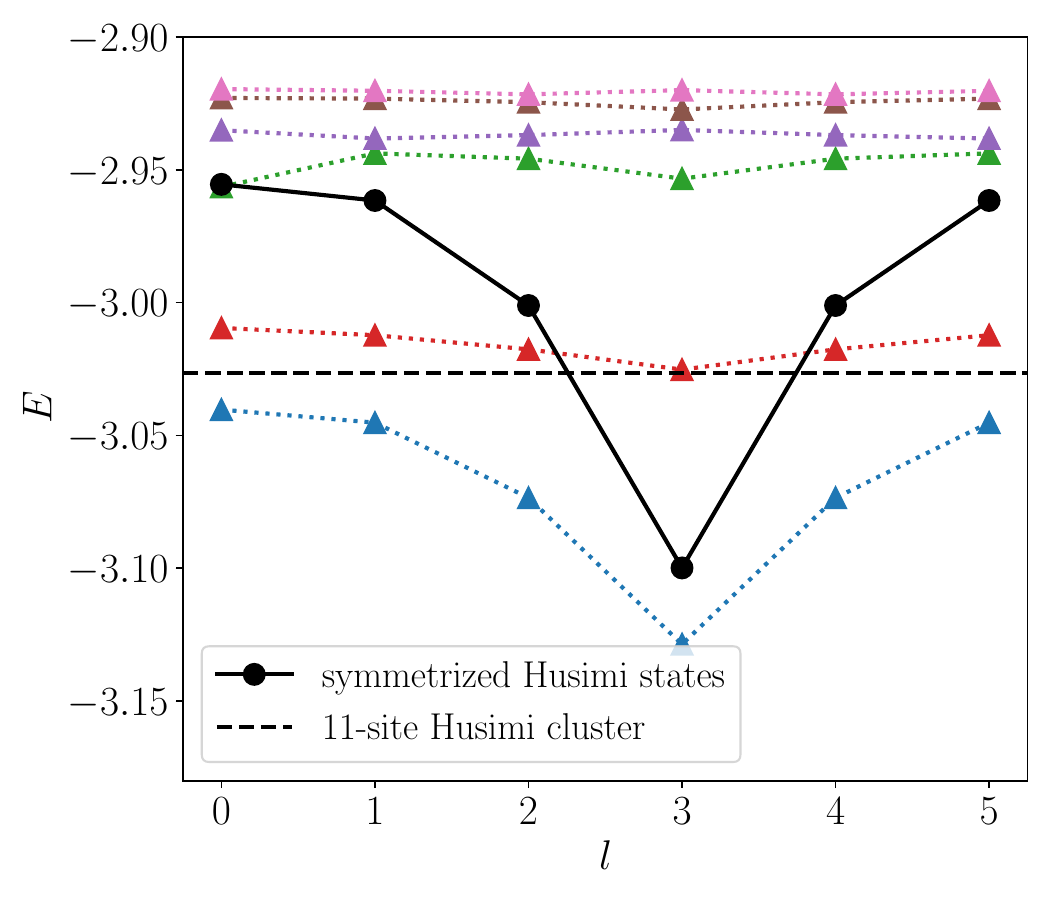}
    \caption{
    \label{fig:ed12_band}Energy levels of the $12$-site sawtooth ring system obtained using ED restricted to each angular momentum quantum number sector, labeled by $l=0, 1, 2, 3, 4, 5$ (colored dashed lines and triangles). The ground state of the system has $l=3$. We also show the GS energy of the $11$-site Husimi clusters, obtained by removing a single triangle from the $12$-site system (dashed horizontal line); as well as the energies of the states obtained by symmetrizing the Husimi cluster states around the ring, again for each value of $l$ (black solid line and circles).} 
\end{figure}

Using ED, we find the GS energy of the system to be $E_{12}=-3.12862$. Owing to the six-fold rotational symmetry, it is possible to partition the Hilbert space into six sectors labeled by the angular momentum quantum number $l=0, 1, 2, 3, 4, 5$, with angular momenta $\mathrm{e}^{\mathrm{i}\frac{2\pi l}{6}}$. We find that the GS has $l=3$. The lower part of the band structure is plotted in Fig.~\ref{fig:ed12_band}. 

Following the same approach used in the main text, we propose a variational GS wavefunction using the lowest-energy states on Husimi cactus clusters of $5$ triangles embedded on the $12$-site system. Eliminating a triangle in the lattice creates a sawtooth chain of length $5$ and open boundary conditions, whose lowest-energy state can be solved for exactly~\cite{kim2023exact, Glittum2026}, and it has energy $E\simeq-3.03$ (horizontal dashed line in Fig.~\ref{fig:ed12_band}). The six states obtained in this way (labeled $|\psi_\mathrm{HC}^{(i)}\rangle, \; i=1,\dots,6$) are not orthogonal; we use them to construct states around the ring $\sum_{i=1}^6\mathrm{e}^{\mathrm{i}\frac{2\pi il}{6}}|\psi_\mathrm{HC}^{(i)}\rangle$ that have different rotational symmetries and different energies, which are plotted in Fig.~\ref{fig:ed12_band}. The state with lowest energy has indeed $l=3$, and we find that it has a $98.5\%$ overlap with the ground state of the $12$-site system. The symmetry of the $12$-sites system is reminiscent of that of the compact-localized wavefunction of the fully polarized system~\cite{Mielke1992ka}, which also has alternating-sign amplitudes.

\end{document}